\documentclass[acmtocl]{acmtrans2m}
\usepackage{amsfonts,amssymb,amsmath}
\usepackage{stmaryrd}
\usepackage{graphics}
\usepackage{pgf}
\usepackage{pgfnodes}
\usepackage{pgfarrows}

\usepackage{tabularx}

\newcommand{\cM}                        {\mathcal{M}}
\newcommand{\cA}                        {\mathcal{A}}
\newcommand{\cE}                        {\mathcal{E}}
\newcommand{\cS}                        {\mathcal{S}}

\newcommand{\sG}                        {\mathsf{G}}
\newcommand{\cC}                        {\mathcal{C}}
\newcommand{\sC}                        {\mathsf{C}}
\newcommand{\cT}                        {\mathcal{T}}
\newcommand{\Q}                        {\mathsf{Q}}
\newcommand{\sA}                        {\mathsf{A}}
\newcommand{\I}                         {\mathcal{I}}
\newcommand{\AP}                        {\mathsf{AP}}
\newcommand{\sL}                        {\mathsf{L}}
\newcommand{\PCTL}                      {{\textsf PCTL}}
\newcommand{\bnfd}                      {\, \talloblong\,}
\newcommand{\cP}                        {\mathcal{P}} 
\newcommand{\verum}                        {\mathsf{tt}}
\newcommand{\falsum}                        {\mathsf{ff}}
\newcommand{\X}                         {\mathsf{X}}
\newcommand{\cU}                        {\mathbin{\mathcal{U}}}
\newcommand{\conj}                      {\mathbin \wedge}
\newcommand{\disj}                      {\mathbin \vee}
\newcommand{\sat}                       {\Vdash}
\newcommand{\ev}                        {\Diamond}
\newcommand{\ACTL}                    {\mathsf{ACTL}}
\newcommand{\NP}                    {\mathsf{NP}}

\newcommand{\DTIME}                    {\mathsf{DTIME}}
\newcommand{\sT}                         {\mathsf{T}}
\newcommand{\mbar}                    {\mathbin{|}}
\newcommand{\ungrp}[1]                {G(#1)}
\newcommand{\lungrp}[1]              {G_\ell(#1)}
\newcommand{\subdistr}[1]             {\mbox{Prob}_{\leq 1}(#1)}
\newcommand{\simd}[1]                 {\preceq_{#1}}
\newcommand{\inj}                   {\mathsf{inj}}
\newcommand{\cB}                 {\mathcal{B}}

\newcommand{\ceil}[1]              {\lceil#1\rceil}

\newcommand{\ic}                            {q_\I}

\newcommand{\Nats}                      {\mathbb{N}}
\newcommand{\Sat}                        {\mathsf{Sat}}
\newcommand{\curr}                        {\mathsf{curr}}
\newcommand{\tmp}                        {\mathsf{tmp}}
\newcommand{\SubForm}               {\mathsf{SLSubForm}}
\newcommand{\PathForm}               {\mathsf{PathForm}}

\newcommand{\MaxProb}                       {\mathsf{MaxProb}}

\newcommand{\id}                         {id}
\newcommand{\cR}                         {\mathcal{R}}
\newcommand{\cK}                         {\mathcal{K}}
\newcommand{\cKex}                         {\mathcal{K}_{\mathsf{ex}}}
\newcommand{\cKab}                         {\mathcal{K}_{\mathsf{ab}}}

\newcommand{\cKcexa}                         {\mathcal{K}_{\mathsf{cex}_1}}
\newcommand{\cKcexb}                         {\mathcal{K}_{\mathsf{cex}_2}}
\newcommand{\cKcexc}                         {\mathcal{K}_{\mathsf{cex}_3}}
\newcommand{\nS}                         {S}
\newcommand{\nT}                         {T}
\newcommand{\R}                          {\mathbin{\cR}}

\newcommand{\st}                        {\mathbin{\mbar}}
\newcommand{\rel}[1]                    {\mathsf{rel}_{#1}}
\newcommand{\post}                      {\mathsf{post}}

\newcommand{\absmdp}[2]              {#1_{#2}}
\newcommand{\amdpcomp}[2]            {#1_{#2}}
\newcommand{\alrel}[1]               {\rel{#1}^{\alpha}}
\newcommand{\gmrel}[1]               {\rel{#1}^{\gamma}}

\newtheorem{theorem}{Theorem}[section]

\newtheorem{proposition}[theorem]{Proposition}
\newtheorem{lemma}[theorem]{Lemma}

\newtheorem{example}[theorem]{Example}

\newenvironment{definition}
  {\begin{trivlist}\item[\hskip\labelsep{\bf Definition:}\small]}{\end{trivlist}}
\newenvironment{notation}
{\begin{trivlist}\item[\hskip\labelsep{\bf Notation:}\small]}{\end{trivlist}}
\newenvironment{remark}
{\begin{trivlist}\item[\hskip\labelsep{\bf Remark:}\small]}{\end{trivlist}}

\newenvironment{claim}
{\begin{trivlist}\item[\hskip\labelsep{\bf Claim:}\small]}{\end{trivlist}}

\title{A Counterexample Guided Abstraction-Refinement Framework for Markov Decision Processes} 
\author{ROHIT CHADHA and MAHESH VISWANATHAN \\ Dept. of Computer Science, University of Illinois at Urbana-Champaign} 

\begin{abstract}
The main challenge in using abstractions effectively, is to construct
a suitable abstraction for the system being verified. One approach
that tries to address this problem is that of {\it counterexample
 guided abstraction-refinement (CEGAR)}, wherein one starts with a
coarse abstraction of the system, and progressively refines it, based
on invalid counterexamples seen in prior model checking runs, until
either an abstraction proves the correctness of the system or a valid
counterexample is generated. While CEGAR has been successfully used in
verifying non-probabilistic systems automatically, CEGAR has not been
applied in the context of probabilistic systems. The main issues that
need to be tackled in order to extend the approach to probabilistic
systems is a suitable notion of ``counterexample'', algorithms to
generate counterexamples, check their validity, and then automatically
refine an abstraction based on an invalid counterexample. In this
paper, we address these issues, and present a CEGAR framework for
Markov Decision Processes.

\end{abstract}
\begin{document}

\maketitle
\section{Introduction}
\label{sec:intro}

Abstraction is an important technique to combat \emph{state space
  explosion}, wherein a smaller, abstract model that conservatively
approximates the behaviors of the original (concrete) system is
verified/model checked. The main challenge in applying this technique
in practice, is in constructing such an abstract model.
\emph{Counterexample guided abstraction-refinement
  (CEGAR)}~\cite{cgjlv00} addresses this problem by constructing
abstractions \emph{automatically} by starting with a coarse
abstraction of the system, and progressively refining it, based on
invalid counterexamples seen in prior model checking runs, until
either an abstraction proves the correctness of the system or a valid
counterexample is generated.

While CEGAR has been successfully used in verifying non-probabilistic
systems automatically, until recently, CEGAR has not been applied in
the context of systems that have probabilistic transitions. In order
to extend this approach to probabilistic systems, one needs to
identify a family of abstract models, develop a suitable notion of
counterexamples, and design algorithms to produce counterexamples from
erroneous abstractions, check their validity in the original system,
and (if needed) automatically refine an abstraction based on an
invalid counterexample. In this paper we address these issues, and
develop a CEGAR framework for systems described as \emph{Markov
  Decision Processes (MDP)}.

Abstractions have been extensively studied in the context of
probabilistic systems with definitions for good abstractions and
specific families of abstractions being identified (see
Section~\ref{sec:related-work}). In this paper,
like~\cite{jl91,Dajjl01,Dajjl02}, we use Markov decision processes to
abstract other Markov decision processes. The abstraction will be
defined by an equivalence relation (of finite index) on the states of
the concrete system. The states of the abstract model will be the
equivalence classes of this relation, and each abstract state will
have transitions corresponding to the transitions of each of the
concrete states in the equivalence class.

Crucial to extending the CEGAR approach to probabilistic systems is to
come up with an appropriate notion of counterexamples. \cite{cjlv02}
have identified a clear set of metrics by which to evaluate any
proposal for counterexamples. Counterexamples must satisfy three
criteria: (a) counterexamples should serve as an ``explanation'' of
why the (abstract) model violates the property, (b) must be rich
enough to explain the violation of a large class of properties, and
(c) must be simple and specific enough to identify bugs, and be
amenable to efficient generation and analysis.

With regards to probabilistic systems there are three compelling
proposals for counterexamples to consider. The first, originally
proposed in~\cite{hk07-1} for DTMCs, is to consider counterexamples to
be a multi-set of executions. This has been extended to
CTMCs~\cite{hk07-2}, and MDPs~\cite{jazzar}. The second is to take
counterexamples to be MDPs with a \emph{tree-like} graph structure, a
notion proposed by~\cite{cjlv02} for non-probabilistic systems and
branching-time logics. The third and final notion, suggested
in~\cite{chatterjee-games,holmanns}, is to view general DTMCs (i.e.,
models without non-determinism) as counterexamples. We show that all
these proposals are expressively inadequate for our purposes. More
precisely, we show that there are systems and properties that do not
admit any counterexamples of the above special forms.

Having demonstrated the absence of counterexamples with special
structure, we take the notion of counterexamples to simply be
``small'' MDPs that violate the property and are simulated by the
abstract model. Formally, a counterexample for a system $\cM$ and
property $\psi_S$ will be a pair $(\cE,\cR)$, where $\cE$ is an MDP
violating the property $\psi_S$ that is \emph{simulated} by $\cM$ via
the relation $\cR$. The simulation relation has rarely been thought of
as being formally part of the counterexample; requiring this addition
does not change the asymptotic complexity of counterexample
generation, since the simulation relation can be computed
efficiently~\cite{bem00}, and for the specific context of CEGAR, they
are merely simple ``injection functions''. However, as we shall point
out, defining counterexamples formally in this manner makes the
technical development of counterexample guided refinement cleaner (and
is, in fact, implicitly assumed to be part of the counterexample, in
the case of non-probabilistic systems).

One crucial property that counterexamples must exhibit is that they be
amenable to efficient generation and analysis~\cite{cjlv02}. We show
that generating the \emph{smallest} counterexample is $\NP$-complete.
Moreover it is unlikely to be efficiently approximable. However, in
spite of these negative results, we show that there is a very simple
polynomial time algorithm that generates a \emph{minimal}
counterexample; a minimal counterexample is a pair $(\cE,\cR)$ such
that if any edge/vertex of $\cE$ is removed, the resulting MDP no
longer violates the property.

Intuitively, a counterexample is valid if the original system can
exhibit the ``behavior'' captured by the counterexample. For
non-probabilistic systems~\cite{cgjlv00,cjlv02}, a valid
counterexamples is not simply one that is simulated by the original
system, even though simulation is the formal concept that expresses the
notion of a system exhibiting a behavior. One requires that the
original system simulate the counterexample, ``in the same manner as
the abstract system''. More precisely, if $\cR$ is the simulation
relation that witnesses $\cE$ being simulated by the abstract system,
then $(\cE,\cR)$ is valid if the original system simulates $\cE$
through a simulation relation that is ``contained within'' $\cR$. This
is one technical reason why we consider the simulation relation to be
part of the concept of a counterexample.  Thus the algorithm for
checking validity is the same as the algorithm for checking
simulations between MDPs~\cite{bem00,zhej07} except that we have to
ensure that the witnessing simulation be ``contained within R''.
However, because of the special nature of counterexamples, better
bounds on the running time of the algorithm can be obtained.

Finally we outline how the abstraction can be automatically refined.
Once again the algorithm is a natural generalization of the refinement
algorithm in the non-probabilistic case, though it is different from
the refinement algorithms proposed
in~\cite{chatterjee-games,holmanns}; detailed comparison can be found
in Section~\ref{sec:related-work}. We also state and prove precisely
what the refinement algorithm achieves.

\subsection{Our Contributions}

We now detail our main technical contributions, roughly in the order
in which they appear in the paper.
\begin{enumerate}
\item For MDPs, we identify safety and liveness fragments of \PCTL .
  Our fragment is syntactically different than that presented
  in~\cite{des99,bkhw05} for DTMCs. Though the two presentations are
  semantically the same for DTMCs, they behave differently for MDPs.
\item We demonstrate the expressive inadequacy of all relevant
  proposals for counterexamples for probabilistic systems, thus
  demonstrating that counterexamples with special graph structures are
  unlikely to be rich enough for the safety fragment of \PCTL .
\item We present formal definitions of counterexamples, their validity
  and consistency, and the notion of good counterexample-guided
  refinements. We distill a precise statement of what the
  CEGAR-approach achieves in a single abstraction-refinement step.
  Thus, we generalize concepts that have been hither-to only defined
  for ``path-like''
  structures~\cite{cgjlv00,cjlv02,hk07-1,hk07-2,jazzar,holmanns} to
  general graph-like structures~\footnote{Even when a counterexample
    is not formally a path, as in~\cite{cjlv02} and~\cite{holmanns},
    it is viewed as a collection of paths and simple cycles, and all
    concepts are defined for the case when the cycles have been
    unrolled a finite number of times.}, and for the first time
  formally articulate, what is accomplished in a single
  abstraction-refinement step.
\item We present algorithmic solutions to all the computational
  problems that arise in the CEGAR loop: we give lower bounds as well
  as upper bounds for counterexample generation, and algorithms to
  check validity and to refine an abstraction. 
\item A sub-logic of our safe-\PCTL, which we call weak safety, does
  indeed admit counterexamples that have a tree-like structure. For
  this case, we present an on-the-fly algorithm to unroll the minimal
  counterexample that we generate and check validity. This algorithm
  may perform better than the algorithm based on checking simulation
  for some examples in practice.
\end{enumerate}
Though our primary contributions are to clarify the definitions and
concepts needed to carry out CEGAR in the context of probabilistic
systems, our effort also sheds light on implicit assumptions made by
the CEGAR approach for non-probabilistic systems. 

\subsection{Outline of the Paper}

The rest of the paper is organized as follows. We recall some
definitions and notations in Section~\ref{sec:prelim}. We also present
safety and liveness fragments of PCTL for MDP's in
Section~\ref{sec:prelim}. We discuss various proposals of
counterexamples for MDP's in Section~\ref{sec:cexam}, and also present
our definition of counterexamples along with algorithmic aspects of
counterexample generation. We recall the definition of abstractions
based on equivalences in Section~\ref{sec:abs}.  We present the
definitions of validity and consistency of abstract counterexamples
and good counterexample-guided refinement, as well as the algorithms
to check validity and refine abstractions in
Section~\ref{sec:refinement}. Finally, related work is discussed in
Section \ref{sec:related-work}.
 
\section{Preliminaries}
\label{sec:prelim}
The paper assumes familiarity with basic probability theory, discrete
time Markov chains, Markov decision processes, and the model checking
of these models against specifications written in PCTL; the background
material can be found in~\cite{marta-book}. This section is primarily
intended to introduce notation, and to introduce and remind the reader
of results that the paper will rely on.

\subsection{Relations and Functions.}
We assume that the reader is familiar with the basic definitions of
relation and functions. We will primarily be interested in binary
relations. We shall use $\cR,\nS,\nT,\ldots$ to range over relations
and $f,g,h,\ldots$ to range over functions.  We introduce here some
notations that will be useful.

Given a set $A$, we shall denote its power-set by $2^A$.  For a finite
set $A$, the number of elements of $A$ shall be denoted by $|A|.$

The identity function on a set $A$ shall be often denoted by $\id_A.$
Given a function $f:A\rightarrow B$ and set $A'\subseteq A$, the
restriction of $f$ to $A'$ shall be denoted by $f|_{A'}.$


For a binary relation $\cR \subseteq A\times B$ we shall often write
$a \R b$ to mean $(a,b)\in \cR$.  Also, given $a\in A$ we shall denote
the set $\{b\in B\st a\R b\}$ by $\cR(a).$ Please note $\cR$ is
uniquely determined by the collection $\{\cR(a)\st a\in A\}.$ A binary
relation $\cR_1\subseteq A\times B$ is said to be {\it finer} than
$\cR_2\subseteq A\times B$ if $\cR_1\subseteq \cR_2.$ The composition
of two binary relations $\cR_1 \subseteq A \times B$ and $\cR_2
\subseteq B \times C$, denoted by $\cR_2\circ \cR_1$, is the relation
$\{(a,c)\st \exists b \in B.\; a\cR_1 b \mbox{ and } b\cR_2 c\}
\subseteq A \times C$.
 

We say that a binary relation $\cR \subseteq A\times B$ is {\it total}
if for all $a\in A$ there is a $b\in B$ such that $a \R b.$ We say
that a binary relation $\cR \subseteq A\times B$ is {\it functional}
if for all $a\in A$ there is at most one $b\in B$ such that $a \R b.$
There is a close correspondence between functions and total,
functional relations: for any function $f: A\to B$, the relation
$\{(a,f(a))\st a\in A\}$ is a total and functional binary relation.
Vice-versa, one can construct a unique function from a given total and
functional binary relation. We shall denote the total and functional
relation given by a function $f$ by $\rel{f}.$
 
A \emph{preorder} on a set $A$ is a binary relation that is reflexive
and transitive. An equivalence relation on a set $A$ is a preorder
which is also symmetric.  The \emph{equivalence class} of an element
$a\in A$ with respect to an equivalence relation $\equiv$, will be
denoted by $[a]_\equiv$; when the equivalence relation $\equiv$ is
clear from the context we will drop the subscript $\equiv$.

\subsection{DTMC and MDP}

\vspace*{0.3cm}\noindent
{\bf Kripke structures.}
A \emph{Kripke structure} over a set of propositions $\AP$, is
formally a tuple $\cK=(\Q,q_\I, \rightarrow, \sL)$ where $\Q$ is a set
of states, $q_\I \in \Q$ is the initial state, $\rightarrow \subseteq
\Q\times \Q$ is the transition function, and $\sL:\Q \to 2^{\AP}$ is a
labeling function that labels each state with the set of propositions
true in it.  DTMC and MDP are generalizations of Kripke structures
where transitions are replaced by probabilistic transitions.

\vspace*{0.3cm}\noindent
{\bf Basic Probability Theory.}
For (finite or countable) set $X$ with $\sigma$-field $2^X$, the
collection all sub-probability measures (i.e., where measure of $X$
$\leq 1$) will be denoted by $\subdistr{X}$. For $\mu \in
\subdistr{X}$, and $A \subseteq X$, $\mu(A)$ denotes the measure of
set $A$.

\vspace*{0.3cm}\noindent
{\bf Discrete Time Markov Chains.}
A \emph{discrete time Markov chain} (DTMC) over a set of propositions
$\AP$, is formally a tuple $\cM=(\Q,q_\I, \delta, \sL)$ where $\Q$ is
a (finite or countable) set of states, $q_\I \in \Q$ is the initial
state, $\delta:\Q \to \subdistr{\Q}$ is the transition function, and
$\sL:\Q \to 2^{\AP}$ is a labeling function that labels each state
with the set of propositions true in it. A DTMC is said to be {\it
  finite} if the set $\Q$ is finite.  Unless otherwise explicitly
stated, DTMC's in this paper will be assumed to be finite.

\vspace*{0.3cm}\noindent
{\bf Markov Decision Processes.}
A \emph{finite Markov decision process} (MDP) over a set of
propositions $\AP$, is formally a tuple $\cM=(\Q,q_\I, \delta, \sL)$
where $\Q, q_\I, \sL$ are as in the case for finite DTMCs, and
$\delta$ maps each state to a \emph{finite} non-empty collection of
sub-probability measures. We will sometimes say that there is no
transition out of $q\in \Q$ if $\delta(q)$ consists of exactly one
sub-probability measure ${\bf 0}$ which assigns $0$ to all states in
$\Q$.  For this paper, we shall assume that for every $q,q'$ and every
$\mu\in\delta(q)$, $\mu(q')$ is a rational number.  From now on, we
will explicitly drop the qualifier ``finite'' for MDP's. In the
presence of a scheduler that resolves nondeterministic choices, a MDP
becomes a (countable) DTMC and a specification is \emph{satisfied} in
an MDP if it is satisfied under all schedulers.

\begin{remark}
  In the presence of memoryless scheduler $\cS$, resulting DTMC is
  {\it bisimilar} to a finite DTMC $\cM^\cS$ which has the same set of
  states as $\cM$, the same initial state and the same labeling
  function, while the transition out of a state $q$ is the one given
  by the memoryless scheduler $\cS$.
\end{remark}

Suppose there are at most $k$ nondeterministic choices from any state
in $\cM$. For some ordering of the nondeterministic choices out of
each states, the \emph{labeled underlying graph} of an MDP is the
directed graph $G = (\Q, \{E_i\}_{i=1}^k)$, where $(q_1,q_2) \in E_i$
iff $\mu(q_2) > 0$, where $\mu$ is the $i$th choice out of $q_1$; we
will denote the labeled underlying graph of $\cM$ by $\lungrp{\cM}$.
The \emph{unlabeled underlying graph} will be $G' = (\Q, \cup_{i=1}^k
E_i)$ and is denoted by $\ungrp{\cM}$. The model checking problem for
MDPs and {\PCTL} is known to be in polynomial time~\cite{Luca}.  The
following notation will be useful.

\begin{notation}
  Given an MDP $\cM=(\Q,q_\I, \delta, \sL)$, a state $q\in \Q$ and a
  transition $\mu\in \delta(q)$, we say that $\post(\mu,q)=\{q'\in
  \Q\st \mu(q')>0.\}$
\end{notation}

\vspace*{0.3cm}\noindent
{\bf Unrolling of a MDP.} 
Given a $\cM=(\Q,q_\I, \delta, \sL)$, natural number $k\geq 0$, and
$q\in \Q$ we shall define an MDP $\cM_k^q=(\Q^q_k,(q,k), \delta_k^q,
\sL_k^q)$ obtained by unrolling the underlying labeled graph of $\cM$
up-to depth $k$.  Formally, $\cM_k^q=(\Q^q_k,(q,k), \delta_k^q,
\sL_k^q)$, the {\it $k$-th unrolling of $\cM$ rooted at $q$} is
defined by induction as follows.
\begin{itemize}
\item $\Q^q_k=\{(q,k)\} \cup (\Q \times \{j\in \Nats\st 0\leq j <
  k\}).$
\item For all $(q',j)\in \Q^q_k$, $\sL((q',j))=\sL(q')$ .
\item For all $(q',j)\in \Q^q_k$, $\delta((q',j))=\{\mu^{j}\st \mu \in
  \delta(q')\} $ where $\mu^j$ is defined as--
          \begin{enumerate}
\item $\mu^{0} (q'')=0$ for all $q''\in \Q^q_k$, and
\item for $0\leq j<k$, $\mu^{j+1} (q'') =\mu(q')$ if $q''=(q',j)$ for
  some $q'\in \Q$ and $0$ otherwise.
\end{enumerate}
\end{itemize}
Please note that the underlying unlabeled graph of $\cM^q_k$ is
(directed) acyclic.

\vspace*{0.3cm}\noindent
{\bf Direct Sum of MDP's.}
Given an MDP's $\cM=(\Q,q_\I, \delta, \sL)$ and $\cM'=(\Q',q'_{\I},
\delta', \sL')$ over the set of propositions $\AP$, let $\Q+\Q'=
\Q\times\{0\} \cup \Q' \times\{1\}$ be the disjoint sum of $\Q'.$ Now,
define $\delta+\delta': \Q+ \Q' \to \subdistr{\Q+\Q'}$ and
$\sL+\sL':\Q+ \Q' \to \AP$ as follows.  For all $q\in \Q$ and $q'\in
\Q'$,
\begin{itemize}  
\item $(\delta+\delta')((q,0))= \{\mu\times\{0\}\st \mu \in
  \delta(q)\}$ and $(\delta+\delta')((q',1))= \{\mu'\times\{1\}\st
  \mu' \in \delta'(q')\}$ where $\mu\times \{0\}$ and $\mu'\times
  \{1\}$ are defined as follows.
\begin{itemize}
\item $\mu\times \{0\} (q_1,0) = \mu(q_1)$ and $\mu\times \{0\}
  (q_1',1) = 0$ for all $q_1 \in \Q_1$ and $q_1' \in \Q_2$.
\item $\mu'\times \{1\} (q_1,0) = 0$ and $\mu'\times \{1\} (q_1',1) =
  \mu'(q_1')$ for all $q_1 \in \Q_1$ and $q_1' \in \Q_2$.
\end{itemize}
\item $(\sL+\sL')(q,0) = \sL(q)$ and $(\sL+\sL')(q',1) =\sL'(q').$
\end{itemize}
Now given $q\in \Q+\Q'$, the MDP $(\cM+\cM')_{q}=(\Q+\Q',q,
\delta+\delta', \sL+\sL')$ is said to be the {\it direct sum} of $\cM$
and $\cM'$ with $q$ as the initial state.

\begin{remark}
  MDP's $\cM=(\Q,q_\I, \delta, \sL)$ and $\cM'=(\Q',q'_{\I}, \delta',
  \sL')$ are said to be {\it disjoint} if $\Q\cap \Q'=\emptyset.$ If
  MDP's $\cM=(\Q,q_\I, \delta, \sL)$ and $\cM'=(\Q',q'_{\I}, \delta',
  \sL')$ are disjoint, then $\Q+\Q'$ can be taken to be the union
  $\Q\cup \Q'$. In such cases, we will confuse $(q,0)$ with $q$,
  $(q',1)$ with $q'$, $\mu \times \{0\}$ with $\mu$ and $\mu'\times
  \{1\}$ with $\mu'$ (with the understanding that $\mu\in \delta(q)$
  takes value $0$ on any $q'\in\Q'$ and $\mu\in \delta(q')$ takes
  value $0$ on any $q\in\Q$).
\end{remark}

\subsection{Simulation}

Given a binary relation $\cR$ on the set of states $\Q$ , a set
$\sA\subseteq \Q$, is said to be {\it $\cR$-closed} if the set
$\cR(\sA)=\{t\,|\,\exists q\in \sA,\, q \R t\}$ is the same as $\sA$.
For two sub-probability measures $\mu,\mu' \in \subdistr{\Q}$, we say
$\mu'$ \emph{simulates} $\mu$ \emph{with respect to} a preorder $\cR$
(denoted as $\mu \simd{\cR} \mu'$) iff for every $\cR$-closed set
$\sA$, $\mu(\sA) \leq \mu'(\sA)$. For an MDP $\cM=(\Q,q_\I, \delta,
\sL)$, a preorder $\cR$ on $\Q$ is said to be a \emph{simulation
  relation} if for every $q\R q'$, we have that $\sL(q)=\sL(q')$ and
for every $\mu \in \delta(q)$ there is a $\mu' \in \delta(q')$ such
that $\mu \simd{\cR} \mu'$.\footnote{It is possible  to  require only
 that
$\sL(q)\subseteq \sL(q')$ instead of $\sL(q)= \sL(q')$ in the definition of simulation. The  results and proofs of the paper could be easily adapted
for this definition. One has to modify the definition of safety and liveness fragments of PCTL appropriately.} We say that $q \preceq q'$ if there is a
simulation relation $\cR$ such that $q \R q'$.

Given an equivalence relation $\equiv$ on the set of states $\Q$, and
two sub-probability measures $\mu,\mu'\in \subdistr{\Q}$ we say that
$\mu$ {\it is equivalent to} $\mu'$ {with respect to} $\equiv$
(denoted as $\mu \approx_{\equiv} \mu'$) iff for every $\equiv$-closed
set $\sA$, $\mu(\sA) = \mu'(\sA)$. For an MDP $\cM=(\Q,q_\I, \delta,
\sL)$, an equivalence $\equiv$ on $\Q$ is said to be a
\emph{bisimulation} if for every $q\R q'$, we have that
$\sL(q)=\sL(q')$ and for every $\mu \in \delta(q)$ there is a $\mu'
\in \delta(q')$ such that $\mu \approx_{\equiv} \mu'$. We say that $q
\approx q'$ if there is a bisimulation relation $\equiv$ such that $q
\equiv q'$.

\begin{remark}
  The ordering on probability measures used in the definition of
  simulation presented in~\cite{jl91,segalalynch,bkhw05} is based on
  \emph{weight functions}. However, the definition presented here, was
  originally proposed in \cite{des} and shown to be
  equivalent~\cite{des,segala}.
\end{remark}

We say that MDP $\cM=(\Q,q_\I, \delta, \sL)$ is simulated by
$\cM'=(\Q',q_\I', \delta', \sL')$ (denoted by $\cM \preceq \cM'$) if
there is a simulation relation $\cR$ on the direct sum of $\cM$ and
$\cM'$ (with any initial state) such that $(q_\I,0) \R (q_\I',1)$. The
MDP $\cM$ is said to be bisimilar to $\cM'$ (denoted by $\cM \approx
\cM'$) if there is a bisimulation relation $\equiv $ on the direct sum
of $\cM$ and $\cM'$ (with any initial state) such that $(q_\I,0)
\equiv (q_\I',1)$.

As an example of simulations, we have that every MDP $\cM=(\Q,q_\I,
\delta, \sL)$ simulates its $k$-th unrolling. Furthermore, we also
have that if $k\leq k'$ then the $k'$-th unrolling simulates the
$k$-th unrolling.
\begin{proposition}
  Given an MDP $\cM$ with initial state $q_\I$ and natural numbers
  $k,k'\geq 0$ such that $k\leq k'$. Let $\cM_k^{q_\I}$ and
  $\cM_{k'}^{q_\I}$ be the $k$-th and $k'$-unrolling of $\cM$ rooted
  at $q_\I$ respectively. Then $\cM_k^{q_\I} \preceq \cM$ and
  $\cM_k^{q_\I} \preceq\cM_{k'}^{q_\I}.$
\end{proposition}

\vspace*{0.3cm}\noindent 
{\bf Simulation between disjoint MDP's.} 
We shall be especially interested in simulation between disjoint MDP's
(in which case we can just take the union of state spaces of the MDP's
as the state space of the direct sum). The simulations will also take
a certain form which we shall call {\it canonical form} for our
purposes. In order to define this precisely, recall that for any set
$A$, $\id_A$ is the identity function on $A$ and that $\rel{{\id_A}}$
is the relation $\{(a,a)\st a \in A\}.$
\begin{definition}
  Given disjoint MDP's $\cM=(\Q,q_\I, \delta,\sL)$ and
  $\cM'=(\Q',q_\I', \delta', \sL')$, we say that a simulation relation
  $\cR\subseteq (\Q+\Q')\times (\Q+\Q')$ on the direct sum of $\Q$ and
  $\Q'$ is in {\it canonical form} if there exists a relation
  $\cR_1\subseteq \Q\times\Q'$ such that $\cR=\rel{\id_\Q} \cup
  \cR_1\cup \rel{\id_{\Q'}}.$
\end{definition}

The following proposition states that any simulation contains a
largest canonical simulation and hence canonical simulations are
sufficient for reasoning about simulation between disjoint MDP's.
\begin{proposition}
  Given disjoint MDP's $\cM=(\Q,q_\I, \delta,\sL)$ and
  $\cM'=(\Q',q_\I', \delta', \sL')$, let $\cR\subseteq (\Q+\Q')\times
  (\Q+\Q')$ be a simulation relation on the direct sum of $\Q$ and
  $\Q'$. Let $\cR_1=\cR\cap (\Q\times \Q').$ Then the relation
  $\cR_0=\rel{\id_\Q} \cup \cR_1\cup \rel{\id_{\Q'}}$ is a simulation
  relation.
\end{proposition}
\begin{proof}
  Clearly $\cR_0$ is reflexive and transitive. Fix $q\in \Q$ and
  $q'\in \Q' $ such that $q\R_0 q'$.  Please note that by definition
  $q\R q'$. Hence, $\sL(q)=\sL(q')$.  We need to show that for any
  $\mu \in \delta(q)$ there is a $\mu_1\in \delta'(q')$ such that $\mu
  \preceq_{\cR_0} \mu_1.$ Since $\cR$ is a simulation relation there
  is a a $\mu'\in \delta'(q')$ such that $\mu \preceq_{\cR} \mu'.$ Fix
  one such $\mu'.$ We claim that $\mu \preceq_{\cR_0} \mu'$ also.
  
  We need to show that for any $\cR_0$-closed set $\Q_0\subseteq
  \Q\cup \Q'$, we have that $\mu(\Q_0)\leq \mu'(\Q_0).$ Now let
  $\Q_1=\Q_0\cap \Q$ and $\Q_2=\Q_0\cap \Q'.$ We have that
  $\mu(\Q_0)=\mu(\Q_1)$ and $\mu'(\Q_0)=\mu'(\Q_2).$ Thus, we need to
  show that $\mu(\Q_1)\leq \mu'(\Q_2).$
  
  Now, consider the set $\cR(\Q_1)=\{q_b\in \Q\cup \Q'\st \exists
  q_a\in \Q_1 \textrm{ s.t. } q_a\R q_b\}.$ Now since $\cR$ is a
  preorder, $\cR(\Q_1)$ is $\cR$-closed and $\Q_1\subseteq
  \cR(\Q_1).$ From $\Q_1\subseteq \cR(\Q_1)$, we can conclude that
  $\mu(\Q_1)\leq \mu(\cR(\Q_1)).$ Also, since $\cR(\Q_1)$ is
  $\cR$-closed and $\mu \preceq_{\cR} \mu'$ we have that
  $\mu(\cR(\Q_1))\leq \mu'(\cR(\Q_1)).$ Hence, we get that
  $\mu(\Q_1)\leq \mu'(\cR(\Q_1)).$ Now, please note that
  $\mu'(\cR(\Q_1))=\mu'(\cR(\Q_1)\cap \Q')$. Hence, the result will
  follow if we can show that $\cR(\Q_1)\cap \Q' \subseteq \Q_2.$
  
  Pick $q_b\in \cR(\Q_1)\cap \Q'.$ We have by definition that $q_b\in
  \Q'$ and there exists $q_a\in \Q_1$ such that $q_a\R q_b. $ Now,
  please note that as $\Q_1\subseteq \Q$, we get $q_a\R_0 q_b$ (by
  definition of $\cR_0$). Also as $\Q_1\subseteq \Q_0$, we get that
  $q_a\in \Q_0$. Since $\Q_0$ is a $\cR_0$-closed set, $q_b\in \Q_0$.
  As $q_b\in\Q'$, we get $q_b\in \Q_2$ also. Since $q_b$ was an
  arbitrary element of $\cR(\Q_1)\cap \Q'$, we can conclude that
  $\cR(\Q_1)\cap \Q' \subseteq \Q_2.$ \qed
\end{proof}

\begin{notation}
  In order to avoid clutter, we shall often denote a simulation
  $\rel{\id_\Q} \cup \cR_1\cup \rel{\id_{\Q'}}$ in the canonical form
  by just $\cR_1$ as in the following proposition. Further, if
  $\cR\subseteq \Q\times\Q'$ is a canonical simulation, then we say
  that any set $A\subseteq \Q \cup \Q'$ is $\cR$-closed iff it is
  $\rel{\id_\Q} \cup \cR \cup \rel{\id_{\Q'}}$-closed.
\end{notation}

\begin{proposition}
  Given pairwise disjoint MDP's $\cM_0=(\Q_0,q_0, \delta_0,\sL_0)$,
  $\cM_1=(\Q_1,q_1, \delta_1,\sL_1)$ and $\cM_2=(\Q_2,q_2,
  \delta_2,\sL_2)$, if $\cR_{01}\subseteq \Q_0\times \Q_1$ and
  $\cR_{12}\subseteq\Q_1\times \Q_2$ are canonical simulations then
  the relation $\cR_{02}=\cR_{12}\circ \cR_{01}\subseteq
  \Q_0\times\Q_2$ is a canonical simulation.
\end{proposition}

\subsection{\PCTL-safety and \PCTL-liveness.}

We define a fragment of {\PCTL} which we call the \emph{safety
  fragment}.  The \emph{safety fragment} of {\PCTL} (over a set of
propositions $\AP$) is defined in conjunction with the \emph{liveness
  fragment} as follows.
\[
{
\begin{array}{l}
\psi_S := \verum \bnfd \falsum \bnfd P \bnfd (\neg P)  \bnfd 
          (\psi_S \conj \psi_S) \bnfd (\psi_S \disj \psi_S) \bnfd 
          \cP_{\triangleleft p}  (\X\, \psi_L ) \bnfd 
          \cP_{\triangleleft p}  (\psi_L\, \cU\, \psi_L )\\
\psi_L := \verum \bnfd \falsum \bnfd P \bnfd (\neg P)  \bnfd 
          (\psi_L \conj \psi_L) \bnfd (\psi_L \disj \psi_L) \bnfd 
          (\neg \cP_{\triangleleft p}  (\X\, \psi_L )) \bnfd 
          (\neg \cP_{\triangleleft p}  (\psi_L\, \cU\, \psi_L ))

\end{array}
}\] where $P \in \AP$, $p \in [0,1]$ is a rational number and
$\triangleleft \in \{<, \leq\}$.  Given a MDP $\cM$ and a state $q$ of
$\cM$, we say $q\sat_\cM \psi$ if $q$ satisfies the formula $\psi.$ We
shall drop $\cM$ when clear from the context. We shall say that
$\cM\sat \psi$ if the initial state of $\cM$ satisfies the formula. 
  
Note that for any safety formula $\psi_S$ there exists a liveness
formula $\psi_L$ such that for state $q$ of a MDP $\cM$, $q
\sat_\cM\psi_S$ iff $q \not\sat_\cM \psi_L$.  Restricting
$\triangleleft$ to be $\leq$ in the above grammar, yields the strict
liveness and weak safety fragments of {\PCTL}. Finally recall that
$\ev \psi$ is an abbreviation for $\verum\, \cU\, \psi$.

There is a close correspondence between simulation and the liveness
and safety fragments of {\PCTL}--- simulation preserves liveness and
reflects safety.

\begin{lemma} 
\label{thm:lem-pctl-safety}
  Let $\cM=(\Q,q_\I, \delta, \sL)$ be an MDP. For any states $q,q' \in
  \Q$, $q \preceq q'$ implies that for every liveness formula
  $\psi_L$, if $q\sat_{\cM} \psi_L$ then $q'\sat_{\cM} \psi_L$ and
  that for every safety formula $\psi_S$, if $q'\sat_{\cM} \psi_S$
  then $q\sat_{\cM} \psi_S$.
\end{lemma}

\begin{proof} 
  The proof is by induction on the length of the safety and liveness
  formulas.  We discuss the case when $\psi_S$ is of the form
  $\Pr_{\triangleleft p } (\psi_{L_1}\cU \psi_{L_2}).$ Assume that
  $q'\sat_{\cM} \psi_S$. We need to show that $q\sat_{\cM} \psi_S$.
  Now if $q\not\sat_{\cM} \psi_{L_1}$ then (by induction)
  $q'\not\sat_{\cM} \psi_{L_1}$. There are two possibilities to
  consider. If $q'\sat_{\cM} \psi_{L_2}$ then $p$ must be $1$ and
  $\triangleleft$ must be $\leq$, and so trivially $q \sat_{\cM}
  \Pr_{\leq 1} (\psi_{L_1}\cU \psi_{L_2})$. On the other hand, if
  $q'\not\sat_{\cM} \psi_{L_2}$ then by induction $q \not\sat_{\cM}
  \psi_{L_2}$ and so $q\sat_{\cM} \Pr_{\triangleleft p }
  (\psi_{L_1}\cU \psi_{L_2}).$
  
  Now let us consider the case when $q\sat_{\cM} \psi_{L_1}$. Let
  $\cR\subseteq \Q\times \Q$ be a simulation relation such that $q\R
  q'$. Now, let $\Q_0\subset \Q$ be the set $\{q_0\in \Q \st q_0
  \sat_{\cM} \psi_{L_1}\disj \psi_{L_2}\}.$ Clearly $q,q'\in \Q_0$.
  Let $\delta_0$ be the restriction of $\delta$ on $Q_0$. That is
  $\delta_0(q)=\{\mu_{|\Q_0}\;|\; \mu\in \delta(q)\}$. Pick a new
  label $P_{\psi_{L_2}}$ and for each each $q_0\in Q_0$ let
  $\sL_0(q_0)=\{P_{\psi_{L_2}}\}$ if $q_0\sat_{\cM} \psi_{L_2}$ and
  $\emptyset$ otherwise. Consider the MDP $\cM_0=(\Q_0,q, \delta_0,
  \sL_0)$. It is easy to see that for any $q_0\in \Q_0$,
  $q_0\sat_{\cM} \psi_S$ iff $q_0\sat_{\cM_0} \Pr_{\triangleleft p }
  (\ev P_{\psi_{L_2}})$.
  
  Let $\cR_0$ be the restriction of $\cR$ to $\Q_0$, {\it i.e.},
  $\cR_0= \cR\cap (Q_0\times Q_0).$ We first show that $\cR_0$ is a
  simulation relation on $\cM_0$ because of the following
  observations.
\begin{enumerate}
\item Reflexivity and transitivity of $\cR_0$ follows from reflexivity
  and transitivity of $\cR$.
\item We claim that if $A\subseteq \cR_0$ is $\cR_0$-closed then $A$
  must also be $\cR$-closed. The proof is by contradiction. Assume
  that there is a $q_1\in \Q\setminus \Q_0$ such that $q_1\in \cR(A).$
  Now, pick $q_0\in A$ such that $q_0\R q_1$.  By construction, either
  $q_0\sat_\cM\psi_{L_1}$ or $q_0\sat_\cM\psi_{L_2}$.  Since $\cR$ is
  a simulation, we get by induction that $q_1\sat_\cM\psi_{L_1}$ or
  $q_1\sat_\cM\psi_{L_2}$. This contradicts $q_1\not\in \Q_0.$
  
  From the above claim it is easy to see that if $\mu\simd{\cR}\mu'$
  then $\mu_{|\Q_0}\simd{\cR_0}\mu'_{|\Q_0}.$ Now, let $q_0 \R_0 q_0'$
  and pick $\mu_0\in \delta_0(q_0)$. We have by definition that $q_0
  \R q_0'$ and there is a $\mu\in \delta(q_0)$ such that
  $\mu_{|Q_0}=\mu_0.$ Since $\cR$ is a simulation there is a $\mu' \in
  \delta(q_0')$ such that $\mu\simd{\cR}\mu'$. We get by the above
  observation, $\mu'_{|\Q_0}\in \delta_0(q_0')$ and
  $\mu_0\simd{\cR_0}\mu'_{|\Q_0}.$
  
\item Similarly we can show that if $q_0\R_0 q_0'$ then
  $\sL_0(q_0)=\sL_0(q_0').$
\end{enumerate}

We have by definition $q \R_0 q'.$ Now, please note we have that
$q'\sat_{\cM_0} \Pr_{\triangleleft p } (\ev P_{\psi_{L_2}})$.  Since
$\ev P_{\psi_{L_2}}$ is a simple reachability formula and $q \R_0 q'$,
results of \cite{jl91} imply that $q\sat_{\cM_0} \Pr_{\triangleleft p
} (\ev P_{\psi_{L_2}})$.  Hence, we get $q\sat_{\cM} \psi_S$.  \qed
\end{proof}

\begin{remark}
\hspace*{1cm}\\
\vspace*{-0.5cm}
\begin{enumerate}
\item The fragment presented here is syntactically different than the
  safety and liveness fragments presented in~\cite{des99,bkhw05} for
  DTMCs; the two presentations have the same semantics for DTMCs, but
  behave differently for MDPs. As far as we know, the safety fragment
  of {\PCTL} for MDPs has not been discussed previously in the
  literature.
  
\item Please note that, unlike the case of DTMCs~\cite{des99,bkhw05},
  logical simulation does not characterize simulation for MDP's. One
  can recover the correspondence between logical simulation and
  simulation, if each non-deterministic choice is labeled uniquely and
  the logic allows one to refer to the label of
  transitions~\cite{radha}.
\end{enumerate}
\end{remark}

\section{Counterexamples}
\label{sec:cexam}

What is a counterexample? \cite{cjlv02} say that
counterexamples must (a) serve as an ``explanation'' of why the
(abstract) model violates the property, (b) must be rich enough to
explain the violation of a large class of properties, and (c) must be
simple and specific enough to identify bugs, and be amenable to
efficient generation and analysis.

In this section, we discuss three relevant proposals for
counterexamples. The first one is due to~\cite{hk07-1}, who present
a notion of counterexamples for DTMCs. This has been recently extended
to MDP's by~\cite{jazzar}.  The second proposal for counterexamples
was suggested in the context of non-probabilistic systems and
branching time properties~\cite{cjlv02}. Finally the third one has
been recently suggested by~\cite{chatterjee-games,holmanns} for MDPs.
We examine the all these proposals in order and identify why each one
of them is inadequate for our purposes. We then present the definition
of counterexamples that we consider in this paper.

\subsection{Set of Traces as Counterexamples}
\label{sec:path-cntrex}

The problem of defining a notion of counterexamples for probabilistic
systems was first considered in ~\cite{hk07-1}. Han and Katoen present
a notion of counterexamples for DTMCs and define a counterexample to
be a {\it finite} set of executions such that the measure of the set
is greater than some threshold (they consider weak-safety formulas
only). The problem to compute the smallest set of executions is
intractable, and Han and Katoen present algorithms to generate such a
set of executions. This definition has recently been extended for MDPs
in~\cite{jazzar}.


\begin{figure}[t]
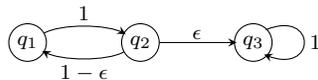

\begin{center}
\begin{pgfpicture}{0cm}{-0.2cm}{4cm}{0.5cm}
\footnotesize
\pgfnodecircle{nA}[stroke]{\pgfxy(0.25,0.25)}{0.25cm}
\pgfnodecircle{nB}[stroke]{\pgfxy(1.75,0.25)}{0.25cm}
\pgfnodecircle{nC}[stroke]{\pgfxy(3.25,0.25)}{0.25cm}
\pgfputat{\pgfnodecenter{nA}}{\pgfbox[center,center]{$q_1$}}
\pgfputat{\pgfnodecenter{nB}}{\pgfbox[center,center]{$q_2$}}
\pgfputat{\pgfnodecenter{nC}}{\pgfbox[center,center]{$q_3$}}
\pgfsetendarrow{\pgfarrowsingle}
\pgfxycurve(0.47,0.375)(0.69,0.5)(1.31,0.5)(1.53,0.375)
\pgfxycurve(1.53,0.125)(1.31,0)(0.69,0)(0.47,0.125)
\pgfnodeconnline{nB}{nC}
\pgfxycurve(3.42,0.42)(4.1,0.7)(4.1,-0.2)(3.42,0.08)
\pgfputat{\pgfxy(1,0.55)}{\pgfbox[center,bottom]{1}}
\pgfputat{\pgfxy(1,-0.05)}{\pgfbox[center,top]{$1-\epsilon$}}
\pgfputat{\pgfxy(2.5,0.3)}{\pgfbox[center,bottom]{$\epsilon$}}
\pgfputat{\pgfxy(4,0.25)}{\pgfbox[left,center]{1}}
\end{pgfpicture}
\end{center}
\caption{DTMC with a large set of counterexample executions}
\label{fig:dtmc-many-paths}
\end{figure}

The proposal to consider a set of executions as a counterexample for
probabilistic systems has a few drawbacks. Consider the DTMC shown in
Figure~\ref{fig:dtmc-many-paths}, where proposition $P$ is true only
in state $q_3$ and $q_1$ is the initial state. Let $\psi_S=\cP_{\leq
  1- \delta}(\ev P)$. The Markov chain violates property $\psi_S$ for
all values of $\delta >0$. However, one can show that the smallest set
of counterexamples is large due to the following observations.
\begin{itemize}
\item Any execution, starting from $q_1$, reaching $q_3$ is of the
  form $(q_1q_2)^kq_3$ with measure $(1-\epsilon)^{k-1}
  \epsilon$. Thus the measure of the set $Exec=\{(q_1q_2)^k q_3\,|\,
  k\leq n\}$ is $1-(1-\epsilon)^n$, and the set $Exec$ has size $O(n^2).$
\item Thus the smallest set of examples that witnesses the violation
  of $\psi_S$ has at least $r=\frac{\log \delta}{\log (1-\epsilon)}$
  elements and the number of nodes in this set is $O(r^2)$.
\end{itemize}
$r$ can be very large (for example, take $\epsilon = \frac{1}{2}$ and
$\delta=\frac{1}{2^{2^n}}$). In such circumstances, it is unclear
whether such a set of executions can serve as a comprehensible
explanation for why the system violates the property $\psi_S$.
Further, this DTMC also violates the property $\psi_S=\cP_{<1}(\ev
P)$. However, there is no finite set of executions that witnesses this
violation. Such properties are not considered
in~\cite{hk07-1,jazzar}.
  
\subsection{Tree-like Counterexamples}
\label{sec:tree-like-cntrex}

In the context of non-probabilistic systems and branching-time
properties, \cite{cjlv02} suggest that counterexamples
should be ``tree-like''. The reason to consider this proposal
carefully is because probabilistic logics like {\PCTL} are closely
related to branching-time logics like $\mathsf{CTL}$. \emph{Tree-like
  counterexamples} for a Kripke structure $\cK$ and property $\varphi$
are defined to be a Kripke structure $\cE$ such that (a) $\cE$
violates the property $\varphi$, (b) $\cE$ is simulated by $\cK$, and
(c) the underlying graph of $\cE$ is \emph{tree-like}, i.e., (i) every
non-trivial maximal strongly connected component is a cycle, and (ii)
the graph of maximal strongly connected components forms a tree.
\cite{cjlv02} argue that this is the appropriate notion of
counterexamples because tree-like counterexamples are easy to
comprehend. Moreover, they show that for any Kripke structure $\cK$
that violates an $\mathsf{ACTL^*}$ formula $\varphi$, there is a
tree-like counterexample $\cE$.

The notion of tree-like counterexamples can be naturally extended to
the case of MDPs. Formally, a \emph{tree-like counterexample} for a
MDP $\cM$ and property $\psi_S$ will be a (disjoint) MDP $\cE$ such
that the unlabeled underlying graph $\ungrp{\cE}$ is tree-like, $\cE$
violates property $\psi_S$ and is simulated by $\cM$. However,
surprisingly, unlike the case for Kripke structures and
$\mathsf{ACTL^*}$, the family of tree-like counterexamples is not rich
enough.

\begin{figure}[t]
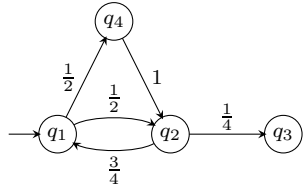

\begin{center}
\begin{pgfpicture}{0cm}{-0.2cm}{4cm}{2cm}
\footnotesize
\pgfnodecircle{nA}[stroke]{\pgfxy(0.25,0.25)}{0.25cm}
\pgfnodecircle{nB}[stroke]{\pgfxy(1.75,0.25)}{0.25cm}
\pgfnodecircle{nC}[stroke]{\pgfxy(3.25,0.25)}{0.25cm}
\pgfnodecircle{nD}[stroke]{\pgfxy(1,1.75)}{0.25cm}
\pgfputat{\pgfnodecenter{nA}}{\pgfbox[center,center]{$q_1$}}
\pgfputat{\pgfnodecenter{nB}}{\pgfbox[center,center]{$q_2$}}
\pgfputat{\pgfnodecenter{nC}}{\pgfbox[center,center]{$q_3$}}
\pgfputat{\pgfnodecenter{nD}}{\pgfbox[center,center]{$q_4$}}
\pgfsetendarrow{\pgfarrowsingle}
\pgfxyline(-0.4,0.25)(0,0.25)
\pgfxycurve(0.47,0.375)(0.69,0.5)(1.31,0.5)(1.53,0.375)
\pgfxycurve(1.53,0.125)(1.31,0)(0.69,0)(0.47,0.125)
\pgfnodeconnline{nB}{nC}
\pgfnodeconnline{nA}{nD}
\pgfnodeconnline{nD}{nB}
\pgfputat{\pgfxy(1,0.55)}{\pgfbox[center,bottom]{$\frac{1}{2}$}}
\pgfputat{\pgfxy(1,-0.05)}{\pgfbox[center,top]{$\frac{3}{4}$}}
\pgfputat{\pgfxy(2.5,0.3)}{\pgfbox[center,bottom]{$\frac{1}{4}$}}
\pgfputat{\pgfxy(0.5,1)}{\pgfbox[right,center]{$\frac{1}{2}$}}
\pgfputat{\pgfxy(1.5,1)}{\pgfbox[left,center]{$1$}}
\end{pgfpicture}
\end{center}
\caption{DTMC $\cM_{notree}$: No tree-like counterexamples}
\label{fig:dtmc-no-tree}
\end{figure}

\begin{example}\rm
\label{exam:no-tree-like}
Consider the DTMC $\cM_{notree}$ shown in
Figure~\ref{fig:dtmc-no-tree} with initial state $q_1$,
proposition $P$ being true only in state $q_3$, and propositions
$P_1,P_2$ and $P_4$ being true only is states $q_1,q_2$ and $q_4$
respectively. Consider the formula $\psi_S=\cP_{<1}((P_1\disj P_2
\disj P_4)\cU P)$. Clearly, the DTMC $\cM_{notree}$ violates $\psi_S.$
\end{example}

We will show that there is no tree-like counterexample for
$\cM_{notree}$ and formula $\psi_S$, defined in
Example~\ref{exam:no-tree-like}. We start by showing that there is no
tree-like DTMC counterexample for $\cM_{notree}$ and $\psi_S$.

\begin{proposition}
\label{prop:notree}
Consider the DTMC $\cM_{notree}$ and safety formula $\psi_S$ defined
in Example~\ref{exam:no-tree-like}. If $\cT=(\Q,q_\I,\delta,\sL)$ is a
DTMC (disjoint from $\cM_{notree}$) such that $\cT\preceq
\cM_{notree}$ and $\cT\not\sat \psi_S$ then $\cT$ is not tree-like.
\end{proposition}

\begin{proof}
  Assume, by way of contradiction, that $\cT$ is tree-like.  Let
  $\Q_0=\{q_1,q_2,q_3,q_4\}$ and for each $1\leq i\leq 4$, let
  $\mu_{q_i}$ denote the transition out of $q_i$ in $\cM_{notree}.$
  
  For each $q\in \Q$, let $\mu_{q}$ denote the transition out of $q$
  in $\cT.$ Let $\cR\subseteq \Q\times \Q_0$ be a canonical simulation
  that witnesses the fact that $\cT\preceq\cM_{notree}$. We have by
  definition, $q_\I\R q_1.$ Please note that since $\cT$ violates
  $\psi_S$ the measure of all paths of $\cT$ starting with $q_\I$ and
  satisfying $(P_1\disj P_2\disj P_4) \cU P$ is $1$.
  
  As $\cT$ is tree-like, any non-trivial strongly connected components
  of $\ungrp{\cT}$ is a cycle and $\sG$, the graph of the strongly
  connected components (trivial or non-trivial) of $\ungrp{\cT}$ form
  a tree. Without loss of generality, we can assume that the strongly
  connected component that forms the root of $\sG$ contains $q_\I$
  (otherwise we can just consider the DTMC restricted to the states
  reachable from $q_\I$).
  
  Hence, we have that every state in $\Q$ is reachable from $q_\I$
  with non-zero probability. From this and the fact $\cR$ is a
  canonical simulation, we can show that for any state $q\in \Q$ there
  is a $q'\in \Q_0$ such that $q\R q'.$ Also since each state in
  $\Q_0$ is labeled by a unique proposition, it follows that for each
  $q\in\Q$ there is a unique $q'\in \Q_0$ such that $q\R q'$ (in other
  words, $\cR$ is total and functional).
  
  Now, for each $1\leq i\leq 4$, let $\Q_i\subseteq \Q$ be the set
  $\{q\in \Q \st q \R q_i.\}$ By the above observations, we have that
  $Q_i$'s are pairwise disjoint; $\Q=\Q_1\cup \Q_2\cup \Q_3\cup \Q_4$;
  and for each $1\leq i\leq 4$, $\Q_i\cup\{q_i\}$ is a $\cR$-closed
  set.  Since $\cR$ is a canonical simulation, whenever $q\cR q'$, we
  have $\mu_q(Q_i) \leq \mu_{q'}(q_i)$, for each $1\leq i\leq 4$.
  Moreover, we can, in fact, prove the following stronger claim.
\begin{claim}
 $\mu_q(\Q_i)=  \mu_{q'}(q_i)$ for each $q\cR q'$ and $1\leq i\leq 4$.
\end{claim}

\noindent
{\bf Proof of the claim:}
Consider some $q,q'$ such that $q\R q'$.  We proceed by contradiction.
Assume that there is some $i$ such $\mu_q(\Q_i) < \mu_{q'}(q_i).$
Please note that in this case $q'\ne q_3$ (as $\mu_{q_3}(q_i)=0,
\forall 1\leq i\leq 4$).

There are several possible cases (depending $q'$ and $q_i$). We just
discuss the case when $q'$ is $q_4$ and $i$ is $2$. The other cases
are similar. For this case we have that $\mu_q(\Q_2)< 1.$ Also note
that for $j\ne 2$, $\mu_q(\Q_j)\leq \mu_{q_4}(q_j)=0.$ Hence,
$\mu_{q}(\Q)<1.$ Now, pick two new states $q_{new_2}$ and $q_{new_3}$
not occurring in $\Q\cup \Q_0$. Construct a new tree-like DTMC $\cT'$
extending $\cT$ as follows. The states of $\cT'$ are $\Q \cup
\{q_{new_2}, q_{new_3}\}.$ Only proposition $P_2$ is true in
$q_{new_2}$ and only proposition $P$ is true in $q_{new_3}$. The
labeling function for other states remains the same. We extend the
probabilistic transition $\mu_q$ by letting
$\mu_q(q_{new_2})=1-\mu_q(\Q)$ and $\mu_q(q_{new_3})=0$ (transition
probabilities to other states do not get affected). The state
$q_{new_2}$ has a probabilistic transition $\mu_{q_{new_2}}$ such that
$\mu_{q_{new_2}}(q_{new_3})=\frac{1}{4}$ and
$\mu_{q_{new_2}}(\bar{q})=0$ for any $\bar{q}\ne q_{new_3}$.  The
transition probability from ${q_{new_3}}$ to any state is $0.$ For all
other states the transitions remain the same.
 
Now, please note that there is a path $\pi$ from $q_\I$ to $q$ (in
$\cT$ and hence in $\cT'$ also) with non-zero ``measure'' such that
$P_1\disj P_2\disj P_4$ is true at each point in this path.
Furthermore, at each point in this path, $P$ is false.  Consider the
path $\pi'$ in $\cT'$ obtained by extending $\pi$ by $q_{new_2}$
followed by $q_{new_3}.$ Now, by construction $\pi'$ satisfies
$(P_1\disj P_2\disj P_4)\cU P$ and the ``measure'' of this path $>0$
(as $1-\mu_q(\Q)>0$). Now the path $\pi'$ is not present in $\cT$ and
hence the measure of all paths in $\cT'$ that satisfy $(P_1\disj
P_2\disj P_4)\cU P$ is strictly greater than the measure of all paths
in $\cT$ that satisfy $(P_1\disj P_2\disj P_4)\cU P$. The latter
number is $1$ and thus the measure of all paths in $\cT'$ that satisfy
$(P_1\disj P_2\disj P_4)\cU P>1.$ Impossible! \qed({End proof of the
  claim})

We proceed with the proof of the main proposition. Let $\cR_1\subseteq
(\Q\cup\Q_0)\times (\Q\cup\Q_0)$ be the reflexive, symmetric and
transitive closure of $\cR$ (in other words, the smallest equivalence
that contains $\cR$). It is easy to see that the equivalence classes
of $\cR_1$ are exactly $\Q_i\cup \{q_i\}, 1\leq i\leq 4.$ From this
fact and our claim above, we can show that the $\cR_1$ is a
bisimulation.
 
Observe now that each element of $\Q_3\subseteq \Q$ must be a leaf
node of $\sG$, the graph of the strongly connected components of
$\ungrp{\cT}.$ Using this, one can easily show that if $\cT_1$ is the
DTMC obtained from $\cT$ by restricting the state space to
$\Q_1\cup\Q_2\cup\Q_4$, then $\cT_1$ is tree-like. Let $\cM_1$ be the
DTMC obtained from $\cM_{notree}$ by restricting the state space to
$\Q_0\setminus\{q_3\}$ and $\tilde{\Q}= (\Q_1\cup\Q_2\cup\Q_4)\cup
(\Q_0\setminus\{q_3\}).$ It is easy to see that the the equivalence
relation $\cR_2=\cR_1\cap (\tilde{\Q}\times\tilde{\Q})$ is also a
bisimulation.
 
Now, let $\sG_1$ be the graph of strongly connected components of
$\ungrp{\cT_1}.$ Now, fix a strongly connected component of
$\ungrp{\cT_1}$, say $\sC$, that is a leaf node of $\sG_1$. Fix a
state $q$ which is a node of $\sC$. Since $\sG_1$ is tree-like and
$\sC$ is a leaf node, it is easy to see that $\post(\mu_q,q)$ can
contain at most $1$ element. Also, we have that $q\in\Q_1\cup \Q_2\cup
\Q_4.$ Now if $q\in \Q_1$, we have that $q$ (as a state of $\cT_1$) is
bisimilar to $q_1$ (as a state of $\cM_1$). However, this implies that
$\post(\mu_q,q)$ must be at least $2$ as $q_1$ has a non-zero
probability of transitioning to $2$ states labeled by different
propositions. Hence $q\not\in \Q_1.$ If $q\in \Q_2$, then please note
that $\post(\mu_q,q)$ must contain an element in $\Q_1$ which should
also be in $\sC$. By the above observation this is not possible. Hence
$q\not\in \Q_2.$ Similarly, we can show that $q\not\in \Q_4$. Hence
$q\not\in\Q_1\cup \Q_2\cup \Q_4.$ A contradiction.  \qed\end{proof}

We are ready to show that $\cM_{notree}$ has no tree-like counterexamples.

\begin{lemma}
\label{lemma:notree}
Consider the DTMC $\cM_{notree}$ and formula $\psi_S$ defined in
Example~\ref{exam:no-tree-like}. There is no tree-like counterexample
witnessing the fact that $\cM_{notree}$ violates $\psi_S$.
\end{lemma}

\begin{proof}
  First, since $((P_1\disj P_2\disj P_3) \cU P)$ is a simple
  reachability formula, if there is a MDP $\cE\preceq \cM_{notree}$
  which violates $\psi_S$, then there is a memoryless scheduler $S$
  such that $\cE^S$ violates the property \cite{Luca}. Now note that
  if we were to just consider the states of $\cE^S$ reachable from the
  initial state of then $\cE^S$ is also tree-like.  In other words,
  there is a tree-like DTMC that is simulated by $\cM_{notree}$ and
  which violates the property $\psi_S$.  The result now follows from
  Proposition ~\ref{prop:notree}.
\end{proof}

Tree-like graph structures are not rich enough for PCTL-safety.
However, it can be shown that if we restrict our attention to
weak-safety formulas, then we have tree counterexamples. However, such
trees can be very big as they depend on the actual transition
probabilities.

\begin{theorem}
\label{thm:tree-cntr-strict-liveness}
If $\psi_{WS}$ is a weak safety formula and $\cM\not\sat \psi_{WS}$,
then there is a $\cM'$ such that $\ungrp{\cM'}$ is a tree, $\cM'
\preceq \cM$, and $\cM' \not\sat \psi_{WS}$.
\end{theorem} 

\begin{proof}
The result follows from the following two observations.
\begin{itemize}
\item If the underlying graph $\ungrp{\cM_1}$ of a MDP $\cM_1$ is
  acyclic then there is an MDP $\cM_2$ such that $\ungrp{\cM_2}$ is a
  tree and $\cM_1\approx \cM_2.$
\item For any strict liveness formula $\psi_{SL}$, and a state $q\in
  \cM$ if $ q\sat_\cM\psi_{SL}$ then there is a $k$ such that
  $\cM_k^{q}\sat \psi_{SL}$ where $\cM_k^{q}$ is the $k$-th unrolling
  of $\cM$ rooted at $q$.
\end{itemize}
The latter observation can be proved by a straightforward induction on
the structure of strict liveness formulas. We consider the case when
$\psi_{SL}$ is $(\neg \cP_{\leq p} (\psi_{SL_1}\, \cU\, \psi_{SL_2}
))$.
          
Now if $q\sat\neg \cP_{\leq p}( \psi_{SL_1}\, \cU\, \psi_{SL_2}) $
then note that there is a memoryless scheduler $\cS$ \cite{Luca} such
that $q\sat_{\cM^\cS} \neg \cP_{\leq p} (\psi_{SL_1}\, \cU\,
\psi_{SL_2}). $ This implies that there is a finite set of finite
paths of $\cM^\cS$ starting from $q$ such that each path satisfies
$\psi_{SL_1}\, \cU\, \psi_{SL_2}$ and the measure of these paths $> p$
\cite{hk07-1}.  Now, these paths can be arranged in a tree $\sT$ nodes
of which are labeled by the corresponding state of $\cM$. If the state
$q'$ labels a leaf node then we have $q'\sat _\cM \psi_{SL_2}$;
otherwise $q'\sat_\cM\psi_{SL_1}.$

For any state $q'\in \Q$ labeling a node in $\sT$ define
$\mathsf{maxdepth}(q')=\max\{\mathsf{depth}(t)\st t \textrm{ is a node
  of } \sT \textrm{ and }t \textrm{ is labeled by }q'\}. $ If $q'$
labels a node of $\sT$ such that $q'\sat_\cM\psi_{SL_1}$ but
$q'\not\sat_\cM\psi_{SL_2}$ then fix $k_{q'}$ such that
$\cM_{k_{q'}}^{q'}\sat \psi_{SL_1}$ ($k_{q'}$ exists by induction
hypothesis). If $q'$ labels a node of $\sT$ such that
$q'\sat_\cM\psi_{SL_2}$ but $q'\not\sat_\cM\psi_{SL_1}$ then fix
$k_{q'}$ such that $\cM_{k_{q'}}^{q'}\sat \psi_{SL_2}$. If $q'$ labels
a node of $\sT$ such that $q'\sat_\cM\psi_{SL_1}$ and
$q'\sat_\cM\psi_{SL_1}$ then fix $k_{q'}$ such that
$\cM_{k_{q'}}^{q'}\sat \psi_{SL_1}$ and $\cM_{k_{q'}}^{q'}\sat
\psi_{SL_2}$.  Now, let $k=\max\{\mathsf{maxdepth}(q')+k_{q'}\st q'
\textrm{ labels a node of } \sT .\}$ It can now be shown easily that
$\cM_k^{q}\sat\psi_{SL}.$
\end{proof}

\subsection{DTMCs as counterexamples}
\label{sec:no-dtmc}

We now consider the third and final proposal for a notion of
counterexamples that is relevant for MDPs. \cite{chatterjee-games} use
the idea of abstraction-refinement to synthesize winning strategies for
stochastic 2-player games. They abstract the game graph, construct
winning strategies for the abstracted game, and check the validity of
those strategies for the original game. They observe that for
discounted reward objectives and average reward objectives, the
winning strategies are \emph{memoryless}, and so ``counterexamples''
can be thought of as finite-state models without non-determinism
(which is resolved by the strategies constructed).

This idea also used in~\cite{holmanns}. They observe that for
weak-safety formulas of the form $\cP_{\leq p}(\psi_1\cU\psi_2)$ where
$\psi_1$ and $\psi_2$ are propositions (or boolean combinations of
propositions), if an MDP $\cM$ violates the property then there is a
memoryless scheduler $\cS$ such that the DTMC $\cM^\cS$ violates
$\cP_{\leq p}(\psi_1\cU\psi_2)$ (see~\cite{Luca}). Therefore, they
take the pair $(\cS,\cM^\cS)$ to be the counterexample.

Motivated by these proposals and our evidence of the inadequacies of
sets of executions and tree-like systems as counterexamples, we ask
whether DTMCs (or rather purely probabilistic models) could serve as
an appropriate notion for counterexamples of MDPs. We answer this
question in the negative.

\begin{proposition}
There is a MDP $\cM$ and a safety formula $\psi_S$ such that $\cM
\not\sat \psi_S$ but there is no DTMC $\cM'$ that violates $\psi_S$ and is
simulated by $\cM$.
\end{proposition}

\begin{proof}
  The MDP $\cM$ will have three states $q_0,q_1,q_2.$ The transition
  probability from $q_1$ and $q_2$ to any other state is $0$.  There
  will be two transitions out of $q_0$, $\mu_1$ and $\mu_2$, where
  $\mu_1(q_0)=0,\mu_1(q_1)=\frac{3}{4}, \mu_1(q_2)=\frac{1}{4}$ and
  $\mu_2(q_0)=0,\mu_2(q_1)=\frac{1}{4}, \mu_2(q_2)=\frac{3}{4}$.  For
  the labeling function, we pick two distinct propositions $P_1$ and
  $P_2$ and let $\sL(q_0)=\emptyset$, $\sL(q_1)=\{P_1\}$ and
  $\sL(q_2)=\{P_2\}$.  Consider the safety formula $\psi_S=\cP_{<
    \frac{3}{4}} (\X (P_1\conj \neg P_2)) \disj \cP_{< \frac{3}{4}}
  (\X (\neg P_1\conj P_2)).$ Now $\cM$ violates $\psi_S$.
  
  Suppose that $\cM'=(\Q',q_\I,\delta',\sL')$ is a counterexample for
  $\cM$ and $\psi_S$. Then we must have $q_\I\sat \neg \cP_{<
    \frac{3}{4}} (\X (P_1\conj \neg P_2)) \conj \neg \cP_{<
    \frac{3}{4}} (\X (\neg P_1\conj P_2))$. Now, if $\cM'$ is a DTMC,
  $\delta'(q_\I)$ must contain exactly one element $\mu_{q_\I}.$ Also
  since $q_\I\sat\neg \cP_{< \frac{3}{4}} (\X (P_1\conj \neg P_2))$
  there must be a state $q_1'$ such that $P_1\in\sL'(q_1'),
  P_2\not\in\sL'(q_1')$ and $\mu_{q_\I}(q_1')\geq\frac{3}{4}.$
  Similarly, there must also be a state $q_2'$ such that
  $P_2\in\sL'(q_2'), P_1\not\in\sL'(q_2')$ and
  $\mu_{q_\I}(q_2')\geq\frac{3}{4}.$ Now, clearly $q_1'\ne q_2'$ (as
  they do not satisfy the same set of propositions).  However, we have
  that $\mu_{q_\I}(Q')\geq\frac{3}{4}+\frac{3}{4}>1.$ A contradiction.
  \qed
\end{proof}

\subsection{Our Proposal: MDPs as Counterexamples}
\label{sec:def-cntrex}


Counterexamples for MDPs with respect to safe PCTL formulas cannot
have any special structure. We showed that there are examples of MDPs
and properties that do not admit any tree-like counterexample
(Section~\ref{sec:tree-like-cntrex}). We also showed that there are
examples that do not admit collections of executions, or general DTMCs
(i.e., models without nondeterminism) as counterexamples
(Sections~\ref{sec:path-cntrex}, \ref{sec:no-dtmc}). Therefore in our
definition, counterexamples will simply be general MDPs. We will
further require that counterexamples carry a ``proof'' that they are
counterexamples in terms of a canonical simulation which witnesses the
fact the given MDP simulates the counterexample. Although we do not
really need to have this simulation in the definition for discussing
counterexamples (one can always compute a simulation), this slight
extension will prove handy while discussing counterexample guided
refinement. Formally,

\begin{definition}
  For an MDP $\cM=(\Q,q_\I,\delta,\sL)$ and safety property $\psi_S$
  such that $\cM\not\sat \psi_S$, a counterexample is pair $(\cE,\cR)$
  such that $\cE=\{\Q_\cE,q_\cE,\delta_\cE,\sL_\cE\}$ is an MDP
  disjoint from $\cM$, $\cE\not\sat \psi_S$ and $\cR\subseteq
  \Q_\cE\times \Q$ is a canonical simulation.
\end{definition}

For the counterexample to be useful we will require that it be
``small''. Our definition of what it means for a counterexample to be
``small'' will be driven by another requirement outlined by in
\cite{cjlv02}, namely, that it should efficiently generatable. These
issues will be considered next.

\subsection{Computing Counterexamples}
\label{sec:min-cntrex}

Since we want the counterexample to be small, one possibility would be
to consider the smallest counterexample. The size of a counterexample
$(\cE,\cR)$ can be taken to be the sum of sizes of the underlying
labeled graph of $\cE$, the size of the numbers used as probabilities
in $\cE$ and the cardinality of the set $\cR$; the smallest
counterexample is then the one that has the smallest size.  However,
it turns out that computing the smallest counterexample is a
computationally hard problem. This is the formal content of our next
result. For this section, we assume the standard definition of the size
of a {\PCTL} formula.

\begin{notation}
Given a safety formula, $\psi_S$, we denote the size of $\psi_S$
by $|\psi_S|.$
\end{notation}

We now formally define the size of the counterexample.
\begin{definition}
  Let $\cM=(\Q,\ic,\delta,\sL)$ be a MDP. The size of $\cM$, denoted
  as $|\cM|$, is the sum of the size (vertices+edges) of the labeled
  underlying graph $\lungrp{\cM}$ and the total size of the numbers
  $\cup_{q\in \Q}\{\mu(q')\st q'\in \Q , \mu\in \delta(q),
  \mu(q')>0\}.$ The size of a counterexample $(\cE,\cR)$, denoted as
  $|(\cE,\cR)|$, is the sum of the size of $\cE$ and the cardinality
  (number of elements) of the relation $\cR$.
\end{definition}

Please note that any MDP $\cM$ of size $n$ has a counterexample of
size $\leq 2n$ (just take an isomorphic copy of $\cM$ as the
counterexample MDP and take the obvious ``injection'' as the canonical
simulation relation).

\begin{theorem}
\label{thm:npc-minimum-cntrex}
Given an MDP $\cM$, a safety formula $\psi_S$ such that
$\cM\not\sat\psi_S$, and a number $k\leq 2|\cM|$, deciding whether there
is a counterexample $(\cE,\cR)$ of size $\leq k$ is $\NP$-complete.
\end{theorem}
\begin{proof} 
  The problem is in $\NP$ because one can guess a counterexample
  $(\cE,\cR)$ of size $k$ and check if $\cE$ violates $\psi_S$. The
  hardness result is achieved by a reduction from the exact $3$-cover
  problem~\cite{garey} which is formally defined as follows.
\begin{quote}
  Given a set $\X$ such that $|\X|=3q$ and a collection $\cC$ of
  subsets of $\X$ such that for each $C\in \cC$, $|C|=3$, is there an
  \emph{exact $3$-cover} for $\X$. In other words, is there a
  collection of pairwise disjoint sets $\cB\subseteq \cC$ such that
  $\X=\cup_{B\in \cB}B$.
\end{quote}
Before, outlining the proof, it is useful to recall what a
\emph{$3$-cover} (not necessarily exact) for $\X$ is: The collection
$\cB$ is said to be an {\it $3$-cover}, if $\cB\subseteq \cC$ is a
collection (not necessarily disjoint) such that $\X=\cup_{B\in
  \cB}B$.

Note that without loss of generality we can assume that for each $x
\in \X$ there is a $C\in\cC$ such that $x\in C$ (if this is not the
case, we can simply answer no in polynomial time). Note that $|\cB|=q$
for an exact cover.  Also note that $\X$ has an exact $3$-cover
$\cB\subseteq \cC$ iff there is cover $\cB'\subseteq \cC$ such that
$|\cB'|\leq q.$ (Actually no collection $\cB'$ such that $|\cB'| < q$
can cover $\X$, so $\leq$ is mainly a matter of convenience.)

\begin{figure}[t]
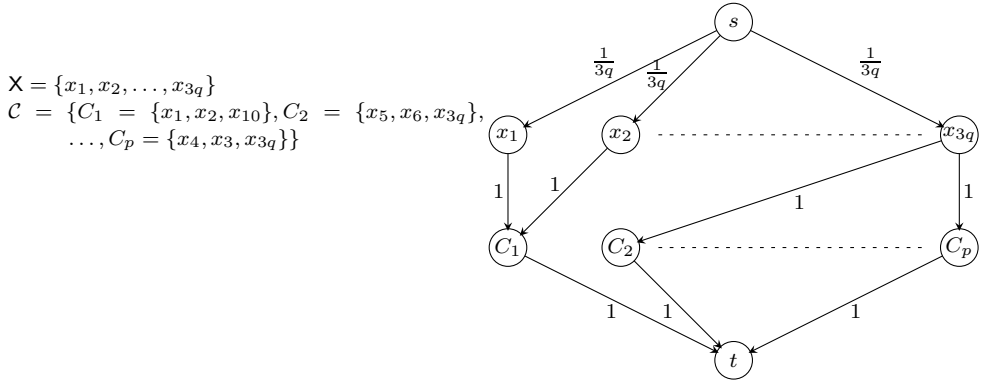

\begin{minipage}[t]{0.5\linewidth}
\vspace*{1cm}
$\X = \{x_1,x_2,\ldots,x_{3q}\}$\\
$\cC = \{C_1 = \{x_1,x_2,x_{10}\},C_2 = \{x_5,x_6,x_{3q}\},$
\hspace*{0.8cm}$\ldots,C_p = \{x_4,x_3,x_{3q}\}\}$
\end{minipage}
\begin{minipage}[t]{0.5\linewidth}
\begin{center}
\begin{pgfpicture}{0cm}{0cm}{6.5cm}{5cm}
\footnotesize
\pgfnodecircle{s}[stroke]{\pgfxy(3.25,4.75)}{0.25cm}
\pgfnodecircle{x1}[stroke]{\pgfxy(0.25,3.25)}{0.25cm}
\pgfnodecircle{x2}[stroke]{\pgfxy(1.75,3.25)}{0.25cm}
\pgfnodecircle{x3q}[stroke]{\pgfxy(6.25,3.25)}{0.25cm}
\pgfnodecircle{c1}[stroke]{\pgfxy(0.25,1.75)}{0.25cm}
\pgfnodecircle{c2}[stroke]{\pgfxy(1.75,1.75)}{0.25cm}
\pgfnodecircle{cp}[stroke]{\pgfxy(6.25,1.75)}{0.25cm}
\pgfnodecircle{t}[stroke]{\pgfxy(3.25,0.25)}{0.25cm}
\pgfputat{\pgfnodecenter{s}}{\pgfbox[center,center]{$s$}}
\pgfputat{\pgfnodecenter{x1}}{\pgfbox[center,center]{$x_1$}}
\pgfputat{\pgfnodecenter{x2}}{\pgfbox[center,center]{$x_2$}}
\pgfputat{\pgfnodecenter{x3q}}{\pgfbox[center,center]{$x_{3q}$}}
\pgfputat{\pgfnodecenter{c1}}{\pgfbox[center,center]{$C_1$}}
\pgfputat{\pgfnodecenter{c2}}{\pgfbox[center,center]{$C_2$}}
\pgfputat{\pgfnodecenter{cp}}{\pgfbox[center,center]{$C_p$}}
\pgfputat{\pgfnodecenter{t}}{\pgfbox[center,center]{$t$}}
\pgfsetendarrow{\pgfarrowsingle}
\pgfnodeconnline{s}{x1}
\pgfnodeconnline{s}{x2}
\pgfnodeconnline{s}{x3q}
\pgfnodeconnline{x1}{c1}
\pgfnodeconnline{x2}{c1}
\pgfnodeconnline{x3q}{c2}
\pgfnodeconnline{x3q}{cp}
\pgfnodeconnline{c1}{t}
\pgfnodeconnline{c2}{t}
\pgfnodeconnline{cp}{t}
\pgfclearendarrow
\pgfsetdash{{0.05cm}{0.1cm}}{0cm}
\pgfxyline(2.25,3.25)(5.75,3.25)
\pgfxyline(2.25,1.75)(5.75,1.75)
\pgfputat{\pgfxy(1.7,4)}{\pgfbox[right,bottom]{$\frac{1}{3q}$}}
\pgfputat{\pgfxy(2.4,4)}{\pgfbox[right,base]{$\frac{1}{3q}$}}
\pgfputat{\pgfxy(4.9,4)}{\pgfbox[left,bottom]{$\frac{1}{3q}$}}
\pgfputat{\pgfxy(0.2,2.5)}{\pgfbox[right,center]{$1$}}
\pgfputat{\pgfxy(0.95,2.5)}{\pgfbox[right,bottom]{$1$}}
\pgfputat{\pgfxy(4.05,2.45)}{\pgfbox[left,top]{$1$}}
\pgfputat{\pgfxy(6.3,2.5)}{\pgfbox[left,center]{$1$}}
\pgfputat{\pgfxy(1.7,1)}{\pgfbox[right,top]{$1$}}
\pgfputat{\pgfxy(2.45,1)}{\pgfbox[right,top]{$1$}}
\pgfputat{\pgfxy(4.8,1)}{\pgfbox[left,top]{$1$}}
\end{pgfpicture}
\end{center}
\end{minipage}
\caption{A problem instance of exact $3$-cover and the constructed MDP}
\label{fig:exact-cover-red}
\end{figure}

The reduction as follows. We first construct an MDP
$\cM=(\Q,q_\I,\delta,\sL)$ as follows. For the set of states, we take
$\Q=\X\cup\cC\cup\{s,t\}$ where $s$ and $t$ are two distinct elements
not in $\X\cup \cC.$ The initial state $q_\I$ is taken to be $s$.
There is one probabilistic transition out of $s$, $\mu_s$, such that
$\mu_s(x)=\frac{1}{3q}$ for each $x\in \X$ and $\mu_s(q)=0$ for all
$q\in \Q\setminus \X$.  From each $x\in\X$, $\delta(x)=\{\mu_{x,C}\st
x\in C,C\in \cC\}$ where $\mu_{x,C}$ assigns probability $1$ to $C$
and $0$ otherwise. For each $C$, there is one probabilistic transition
out of $C$, $\mu_C$, which assigns probability $1$ to $t$ and is $0$
otherwise. There is no transition out of $t$. Finally, the set of
propositions, we will pick a proposition $P_q$ for each $q\in \Q$ and
$P_q$ will be true only in the state $q$. For the safety formula, we
take $\psi_S=\cP_{<1}(\verum \cU P_t).$ Clearly $\cM$ violates
$\psi_S$. The reduction is shown in Figure~\ref{fig:exact-cover-red}.
The result now follows from the following claim.
\begin{claim}
$\X$ has an exact $3$-cover $\cB\subset \cC$ iff there is a counterexample $(\cE,\cR)$ for $\cM$ and
$\psi_S$ of size $\leq 2(2+ 4q)+ 7q+ 3q(1+\ceil{\log 3q}) + 4q$. 
\end{claim}

\noindent
{\bf Proof of the claim:}\\
($\Rightarrow$) Assume that $\cB\subseteq \cC$ is an exact $3$-cover
of $\X$. We have $|\cB|=q$.  Consider an MDP $\cM'$ which is the same
as $\cM$ except that its states are $\{\bar{q}\st q\in \Q\}$ instead
of $\Q$. Now delete all states $\bar{C}$ of $\cM'$ such that $C\notin
\cB$. Let the resulting MDP be called $\cE$ and the set of its states
be denoted by $\Q_\cE.$ Note that the $\ungrp{\cE}$ has $2+4q$ nodes
and $7q$ edges. Furthermore, from the initial state there is a
probabilistic transition which assigns probability $\frac{1}{3q}$ to
each $\{\bar{x}\st x\in \X\}.$ It takes $1+\ceil{\log 3q}$ bits to
represent $1+\ceil{\log 3q}$ ($1$ for the numerator and $\ceil{\log
  3q}$ for the denominator). For each $\bar{x}$ such that $x\in \X$
there is a probabilistic transition which assigns probability $1$ to
$\bar{B}$ where $B\in \cB$ is such that $x\in B$. Finally, from each
$B\in \cB$ there is a probabilistic transition that assigns
probability $1$ to $\bar{t}$. The size of the MDP $\cE$ is seen to be
$2+ 4q+ 7q+ 3q(1+\ceil{\log 3q}) + 4q$.  Now, let $\cR$ be the
relation $\{(\bar{q},q)\st q\in \Q_\cE\}.$ Clearly $(\cE,\cR)$ is a
counterexample and one can easily check that $|\cE,\cR|=2(2+ 4q)+ 7q+
3q(1+\ceil{\log 3q}) + 4q.$

($\Leftarrow$) Assume that there is a counterexample $(\cE,\cR)$ of
size $\leq 2(2+ 4q)+ 7q+ 3q(1+\ceil{\log 3q}) + 4q$.  Thus we have
that $\cE\preceq\cM$, $\cR$ is a canonical simulation and $\cE$
violates $\psi_S.$ Now note that since every node of $\cM$ is labeled
by a unique proposition, $\cR$ is functional. In other words for each
state $q_1$ of $\cE$ is related to at most one state of $\Q.$ Observe
that since the safety formula $\psi_S$ is $\cP_{<1}(\verum \cU P_t)$,
there is a memoryless scheduler $\cS$ such that $\cE^\cS$ violates
$\psi_S.$ Let $\cE^\cS=(\Q_\cE,q_\cE,\delta_\cE,\sL_\cE)$. For each
$q_1\in \Q_\cE$, let $\mu_{q_1}$ denote the unique probabilistic
transition out of $\cE^\cS.$

Note that we have $q_\cE \cR s.$ Consider the set
$\Q_\X=\post(q_\cE,\mu_{q_\cE}).$ Since $\cE$ is simulated by $\cM$;
it follows that each element of $\Q_\X$ must be labeled by some
proposition $P_x$ for some $x\in \X.$ Given $x\in \X$, if
$\Q_x\subseteq \Q_\X$ is the set of states labeled by $P_x$ then we
must have $q\cR x$ for each $q\in \Q_x$. We also have that
$\mu_{q_\cE}(\Q_x)\leq \frac{1}{3q}$ and $\Q_{x_1}\cap
\Q_{x_2}=\emptyset$ for $x_1\ne x_2.$ Now note that the probability of
reaching $P_t$ from $q_\cE$ is $1$ in $\cE^\cS$. Hence it must be the
case that $\mu_{q_\cE}(\Q_\X)=1$ and thus $\Q_x\ne \emptyset$ for any
$x\in \X$. Therefore $|\Q_\X|\geq 3q$ and the total size of the
numbers $\{\mu_{q_\cE}(q)\st \mu_{q_\cE}(q)>0, q\in \Q_\X\}$ is at
least $|\Q_\X|(1+\ceil{\log 3q}).$

Now given $q\in\Q_x$ consider $\post (q,\mu_q).$ Again as the total
probability of reaching $P_t$ is $1$ in $\cE^\cS$, $\post(q,\mu_q)$
cannot be empty. Furthermore, as each $q\in \Q_x$ is simulated by $x$,
it follows that each element $q'\in \post (q,\mu_q)$ must be labeled
by a single $P_B\in \cC$ such that $x\in B$. Let $\Q_\cC=\cup_{q\in
  \Q_\X}\post(q,\mu_q).$ By the above observations it follows that if
we consider the set $\cB=\{B\in \cC \st P_B \textrm{ labels a node in
}\Q_\cC\}$ then $\cB$ covers $\X.$ Also since every state of $\Q_\cC$ is
labeled by a single proposition, we get $|\Q_\cC|\geq |\cB|.$

We can show by similar arguments that for each $q\in \Q_\cC$ the set
$\post(q,\mu_q)$ is non-empty and each node of $\post(q,\mu_q)$ must
be labeled by $P_t.$ Let $\Q_t=\cup_{q\in \Q_\cC}\post(q,\mu_q).$ We
have that $|\Q_t| \geq 1.$

Note that the sets $\Q_\X$, $\Q_t$ and $\Q_\cC$ are pairwise disjoint
and do not contain $q_\cE.$ Hence, the labeled underlying  graph of $\cE$,
$\lungrp{\cE}$, has at least $1+|\Q_\X|+|\Q_\cC|+|\Q_t|$ vertices. As
$\cR$ is functional, $\cR$ also contains at least
$1+|\Q_\X|+|\Q_\cC|+|\Q_t|$ elements. Furthermore, it is easy to see
that the underlying graph has at least $2|\Q_\X|+|\Q_\cC|$ edges; and
the total size of numbers used as probabilities in $\cE$ is at least
$|\Q_\X|(1+\ceil{\log(3t)})+|\Q_\X|+|\Q_\cC|.$ Hence the total size of
$(\cE,R)$ is at least $2(1+|\Q_\X|+|\Q_\cC|+|\Q_t|)+ 2|\Q_\X|+|\Q_\cC|
+ |\Q_\X|(1+\ceil{\log(3t)})+|\Q_\cC|.$ Since $|Q_t|\geq 1$ and
$|\Q_\X|\geq 3q$; the total size is at least $2(2+3q+|\Q_\cC|)+ 6q +
|\Q_\cC| + 3q(1+\ceil{\log(3t)})+3q +|\Q_\cC|.$ By hypothesis, the
total size is $\leq2(2+ 4q)+ 7q+ 3q(1+\ceil{\log 3q}) + 4q$ and we get
that $|\Q_\cC|\leq q.$ But $|\Q_\cC|\geq |\cB|$ and hence $|\cB|\leq q.$
Since $\cB$ is a cover; it follows that $\X$ must have an exact
$3$-cover.  \qed
\end{proof}
Not only is the problem of finding the smallest counterexample
$\NP$-hard, it also in-approximable.
\begin{theorem}
  Given an MDP $\cM$, a safety formula $\psi_S$ and
  $n=|\cM|+|\psi_S|$ such that
  $\cM\not\sat\psi_S$. The smallest counterexample for $\cM$ and
  $\psi_S$ cannot be approximated in polynomial time to within
  $O(2^{\log^{1-\epsilon}n})$ 
  unless $\NP\subseteq \DTIME(n^{poly \log(n)})$.
\end{theorem}
\begin{proof}
  The in-approximability follows from a reduction of the Directed
  Network Steiner Problem~\cite{dk99}.  Directed Network Steiner
  Problem is formally defined as follows.
\begin{quote}
  Given a directed graph $G$, $m$ pairs $\{s_i,t_i\}_{i=1}^m$ a sub-graph
  $G'=(V',E')$ of $G$ satisfies the Steiner condition if $s_i$ has
  path in $G'$ to $t_i$ for all $i$. The Directed Steiner network
  problem asks for a sub-graph $G'$ such that $G'$ satisfies the
  Steiner condition and has the smallest size amongst all subgraphs of
  $G$ which satisfy the Steiner condition.
\end{quote}

It is shown in~\cite{dk99} that the smallest sub-graph cannot be
approximated to within $O(2^{\log^{1-\epsilon}(n_g)})$ where $n_g$ is the sum
$m$+ size (vertices+edges) of $G$ unless $\NP\subseteq \DTIME(n_g^{poly
  \log(n_g)}).$ Also note that since $\epsilon$ is arbitrary the
smallest sub-graph cannot be found to within
$O(2^{\log^{1-\epsilon}(n_g\log (n_g))})$. (Changing $n_g$ to
$n_g\log(n_g)$ does not make a difference $\DTIME(n_g^{poly
  \log(n_g)})$).

We now give the reduction. Given a graph $G =(V,E)$, let $|V|=n_v$,
$|E|=n_e$, $n_g=n_e+n_v$. Recall, that the network steiner problem has
$m$ pairs $(s_i,t_i)$. Let $n_s$ be the number of distinct $s_i
$'s in $(s_i,t_i)$. In other words $n_s$ is the cardinality of the set
$\{s_i\st 1\leq i\leq m\}$. Clearly $n_s\leq n_g$.  Please note that for
the directed network Steiner problem we can assume that $n_g$ is
$O(n_e)$.

We construct a DTMC $\cM$ with states $V \cup \{s\}$, where $s$ is a
new vertex.  $s$ is the initial state of $\cM$ and has a probabilistic
transition $\mu_s$ such that $\mu_s(s_i)= \frac{1}{n_g}$ for each
$1\leq i\leq m$ and $\mu_s(v)=0$ if $v\notin \{s_i|1\leq i\leq m\}.$
Every other state $v$ has a transition $\mu_v$ such that
$\mu_v(v')=\frac{1}{n_g} $ if $(v,v') \in E$; otherwise $\mu_v(v')=0
$.  Finally, we will have as propositions $\{P_v\st v\in
V\cup\{s\}\}$, where the proposition $P_v$ holds at exactly the state
$v$. Since it takes $O(\log(n_g))$ bits to represent $\frac{1}{n_g}$,
the size of DTMC $\cM$ is easily seen to be $n_v+1+ n_e+n_s + (n_e+n_s)
(O(\log(n_g))).$

Consider the safety formula $\psi_S=\psi_{S1} \disj \psi_{S2}$ where
$\psi_{S_1}= \bigvee_{i=1}^{m} \cP_{\leq 0} (\ev (s_i \conj (\neg
\cP_{\leq 0}(\ev t_i))))$ and $\psi_{S_2}=
\bigvee_{i=1}^{n_s}\cP_{\leq 0}(\X\, s_i)$. 

The sum $n_m=|\cM|+|\psi_S|$
is easily seen to be $O(n_g log(n_g)).$

\begin{claim}
\hspace*{1cm}\\
\vspace*{-0.3cm}

\begin{enumerate}
\item If $G$ has a sub-graph $G'=(V',E')$ with $|V_1'| = n_1$ and
  $|E'|= n_2$ such that $G'$ satisfies Steiner condition then $\cM$
  has a counterexample of size $= 2n_1+2+ n_2+n_s + (n_2+n_s)
  (\log(n_g)).$
  
\item If the DTMC $\cM$ has a counterexample $(\cE,\cR)$ such that
  $\lungrp{\cE}$, the underlying labeled graph of $\cE$, has $n_1+1$
  vertices and $n_2+n_s$ edges then the graph $G$ has a sub-graph
  $G'=(V',E')$ with $|V_1'| \leq n_1$ and $|E'|\leq n_2$ such that
  $G'$ satisfies the Steiner condition.  Furthermore, $|(\cE,\cR)|\geq
  2n_1+2+ n_2+n_s + (n_2+n_s) (\log(n_g))$.
\end{enumerate}
\end{claim}

{\bf Proof of the claim:}
\begin{enumerate}
\item First assume that $G$ has a sub-graph $G'=(V',E')$ with less
  than $n_1$ vertices and less than $n_2$ edges such that $G'$
  satisfies the Steiner condition. Consider the DTMC $\cM'$ obtained
  from $\cM$ by restricting the set of states to $V'\cup \{s\}.$ Now
  take an isomorphic copy of $\cM'$ with $\{\bar{v}|v\in V'\} \cup
  {\bar{s}}$ as the set of states and call it $\cE$. Clearly, $\cE$
  violates $\psi_S$ and the relation $\{(\bar{u},u)\st u \in V'\cup
  s\}$ is a canonical simulation of $\cE$ by $\cM$.  Hence $(\cE,\cR)$
  is a counterexample and it is easy to see that $|(\cE,\cR)|\leq
  2n_1+2+ n_2+n_s + (n_2+n_s) (\log(n_g)).$
 
\item Let $\cE=(\Q_\cE,q_\cE,\delta_\cE,\sL_\cE)$ and
  $\lungrp{\cE}=(V',\{E_j'\}^k_{j=1}).$ Note that since each vertex of $\cM$ is
  labeled by a unique proposition and $\cR$ is a canonical simulation,
  $\cR$ must be total and functional (totality is a consequence of the fact 
  that we can remove any nodes of $\cE$ that are
  not reachable from $q_\cE$).  In other words there is a function
  $g:\Q\to V\cup\{s\}$ such that $\cR=\rel{g}$. Again the definition
  of simulation and the construction of DTMC $\cM$ gives us that if
  $(q_1,q_2)\in \cup^k_{j=1}E'_j$, we must have $(g(q_1),g(q_2))\in
  E\cup\{(s,s_i)\;|\;1\leq i\leq m\}$ and $\mu'_{q_1}(q_2)\leq
  \frac{1}{n_g}$ for any probabilistic transition $\mu'_{q_1} \in
  \delta_\cE(q_1)$.  From the latter observation, we get that
  $|(\cE,\cR)|\geq 2n_1+2+ n_2+n_s + (n_2+n_s) (\log(n_g))$.
  
  Consider the equivalence relation $q_1\equiv q_2$ defined on $\Q$ as
  $q_1\equiv q_2$ iff $g(q_1)=g(q_2)$.  Let $[q]$ denote the
  equivalence class of $q$ under $\equiv$. Let $G_2=\{V'',E''\}$ be
  the graph such that $V''$ is the set of equivalence classes under
  the relation $\equiv$ and $([q_1], [q_2])\in E''$ if $(q_1,q_2)\in
  \cup^k_{j=1}E'.$ Please observe first that $G_2$ is isomorphic to a subgraph of
  $(V\cup\{s\}, E\cup \{(s,s_i)\;|\;1\leq i\leq n_s\})$ with the
  function $h([q])= g(q)$ witnessing this graph isomorphism. Please
  note that by the fact that $\cM_1$ violates $\psi_1$, it can be
  easily shown that there is a path from $s_i$ to $t_i$ in $G_2$.
  Also since $\cM_1$ violates $\psi_2$, $G_2$ contains edges $(s,s_i)$
  for each $1\leq i\leq n_s.$ We get by the above observations $G$
  must contain a subgraph $G'=(V',E')$ with paths from $s_i$ to $t_i$
  for all $i$ such that $|V_1'| \leq n_1$ and $|E'|\leq n_2$.  ({\bf
    End proof the claim.})
\end{enumerate}

From the above two observations it easily follows that if $G$ has a
Steiner sub-graph of minimum size $n_{\min_1}$ and $\cM$ has a
counterexample of minimum size $n_{\min_2}$ then
$\frac{n_{\min_2}}{n_{\min_1}}=O(log(n_g)).$ Now assume that there is
a polynomial time algorithm to compute the minimal counterexample
within a factor of $O(2^{\log^{1-\epsilon}(n_m)})$ then this algorithm
produces a counterexample of size $k\leq
O(2^{\log^{1-\epsilon}(n_m)})n_{\min_2}$. Thus the counterexample size
is $\leq O(2^{\log^{1-\epsilon}(n_m)})O(\log(n_g))n_{\min_1}.$ From
the proof of the part $2$ of the above claim it follows that we can
extract from the counterexample in polynomial time a Steiner sub-graph
of $G$ of size $ \leq O(2^{\log^{1-\epsilon}(n_m)}) n_{\min_1}.$ Now
$n_m$ is $O(n_g \log(n_g))$ and thus we have achieved an approximation
within $O(2^{\log^{1-\epsilon}(n_g \log(n_g))})$. The result now
follows.
\qed\end{proof}

\begin{remark}
  A few points about our hardness and inapproximability results are in
  order.
\begin{enumerate}
\item Please note that we did not take the size of the labeling
  function into account.  One can easily modify the proof to take this
  into account.
\item The same reduction also shows a lower bound for the safety
  fragment of $\ACTL^*$ properties as the reduction does not rely on
  any important features of quantitative properties.
\end{enumerate}
\end{remark}

Since finding the smallest counterexamples is computationally hard, we
consider the problem of finding \emph{minimal counterexamples}.
Intuitively, a minimal counterexample has the property that removing
any edge from the labeled underlying graph of the counterexample,
results in an MDP that is not a counterexample. In order to be able to
define this formally, we need the notion of when one MDP is contained
in the other.
\begin{definition}
We say that an MDP $\cM' = (\Q', q_I, \delta', L')$ is contained in an
MDP $\cM = (\Q, q_I, \delta, L)$ if $\Q' \subseteq \Q$, $L'(q') =
L(q')$ for all $q' \in \Q'$, and there is a 1-to-1 function $f:
\delta' \to \delta$ with the following property: For each $q',q'' \in
\Q'$ and $\mu' \in \delta'(q')$, $f(\mu') \in \delta(q')$, either
$\mu'(q'') = f(\mu')(q'')$ or $\mu'(q'') = 0$. We denote this by $\cM'
\subseteq \cM$.
\end{definition}
Observe that if $\cM' \subseteq \cM$ then $\cM' \preceq \cM$.  We
present the definition of minimal counterexamples obtained by
lexicographic ordering on pairs $(\cE,\cR)$.
\begin{definition}
For an MDP $\cM$ and a safety property $\psi_S$, $(\cE,\cR)$ is a minimal
counterexample iff
\begin{itemize}
\item $(\cE,\cR)$ is a counterexample for $\cM$ and $\psi_S$ and
\item If $(\cE_1,\cR_1)$ is also a counterexample for $\cM$ and
  $\psi_S$, then
\begin{itemize}
\item  $\cE_1\subseteq \cE$ implies that $\cE_1=\cE$; and
\item if $\cE_1=\cE$ then  $\cR_1\subseteq \cR_2$ implies that $\cR_1=\cR_2$.
\end{itemize} 
\end{itemize}
\end{definition}

Though finding the smallest counterexample is $\NP$-complete and is
unlikely to be efficiently approximable, there is a very simple
polynomial time algorithm to compute the minimal counterexample. In
fact the counterexample computed by our algorithm is going to be
contained in the original MDP (upto ``renaming'' of states). Before we
proceed, we fix some notation for the rest of the paper.

\begin{notation}
  Given an MDP $\cM=(\Q,q_\I,\delta,\sL)$, for each $q\in \Q$ fix a
  unique element $\bar{q}$ not occurring in $\Q.$ Define an {\it
    isomorphic} MDP
  $\bar{\cM}=(\bar{\Q},\bar{q_\I},\bar{\delta},\bar{\sL})$ as follows.
\begin{itemize}
\item $\bar{\Q}=\{\bar{q}\st q\in \Q\}$.
\item $\bar{\delta}(\bar{q})= \{\bar{\mu}\st\mu \in \delta(q)\}$ where
\begin{itemize}
\item for each $\bar{q}\in \bar{Q}$, $\bar{\mu}(\bar{q})=\mu(q)$.
\end{itemize}
\item $\bar{\sL}(\bar{q})=\sL(q).$
\end{itemize}
\end{notation}

We are ready to give the counterexample generation algorithm.  The
algorithm shown in Figure~\ref{fig:min} clearly computes a minimal
counterexample contained in the original MDP upto ``renaming'' of
states (note that the minimality of $\rel\inj$ is a direct consequence
of the fact that every state in $\cE$ is reachable from initial state
and hence must be simulated by some state in $\cM$). Its running time
is polynomial because model checking problem for MDPs is in
$P$~\cite{Luca}.

\begin{figure}[t]
\begin{center}
\begin{tabular}{|l|}
\hline
\\

Initially $M_{curr}=\bar{\cM}$\\
For each edge $(\bar{q},\bar{q_1}) \in E_i$ in $\lungrp{M_{curr}}$\\
\hspace*{0.5cm}Let $M' $ be the MDP obtained from $M_{curr}$ by setting $\mu(\bar{q_1})
  = 0$, where $\mu$ is the \\
\hspace*{0.8cm} $i$th choice out of $\bar{q}$, in $M_{curr}$\\
\hspace*{0.5cm}If $M' \not\sat \psi_S$ then $M_{curr}={M'}$\\
od $\leftarrow$ End of For loop\\
Let $\cE$ be the MDP obtained from $M_{curr}$ by removing the set of states from 
  $M_{curr}$  \\
  \hspace*{0.3cm}  which are not  reachable in the underlying unlabeled graph of $M_{curr}$\\
  If $\Q_\cE$ is the set of states of $\cE$, then let $\rel\inj=\{(\bar{q},q)\st \bar{q}\in \Q_\cE\}$\\  

return($\cE,\rel\inj$)\\
\\
\hline
\end{tabular}
\end{center}
\caption{Algorithm for computing the minimal counterexample}
\label{fig:min}
\end{figure}

\begin{theorem}
\label{thm:min}
Given an MDP $\cM$ and a safety formula $\psi_S$ such that
$\cM\not\sat\psi_S$, the algorithm in Figure~\ref{fig:min} computes a
minimal counterexample and runs in time polynomial in the size of
$\cM$ and $\psi_S$.
\end{theorem}

Please note that for safety properties of the form $\cP_{\leq
  p}(\psi_S\cU\psi'_S)$ and $\cP_{<p}(\psi_S\cU\psi'_S)$ where
$\psi_S,\psi'_S$ are boolean combinations of propositions, if
$(\cE,\rel\inj)$ is the counterexample generated by Figure
\ref{fig:min} then $\cE$ must be a DTMC.  This is because if $\cE$
violates such a property then there is a memoryless scheduler $\cS$
such that $\cE^\cS$ violates the same property (see \cite{Luca}). For
such properties; model-checking algorithm also computes the memoryless
scheduler witnessing the violation. Thus for such properties, one
could initialize $M_{curr}$ to be $\bar{\cM}^{\cS_1}$ where $\cS_1$ is
the memoryless scheduler generated when $\bar{\cM}$ is model-checked
for violation of the given safety property.

The counterexample returned by the algorithm in Figure~\ref{fig:min},
clearly depends on the order in which edges of $\lungrp{\bar{\cM}}$ are
considered. An important research question is to discover heuristics
for this ordering, based on the property and $\cM$.

\section{Abstractions}
\label{sec:abs} 

Usually, in counterexample guided abstraction refinement framework,
the abstract model is defined with the help of an equivalence relation
on the states of the system \cite{cgjlv00}. Informally, the
construction for non-probabilistic systems proceeds as follows. Given
a Kripke structure $\cK=(\Q,q_\I,\rightarrow,\sL)$ and equivalence
relation $\equiv$ on $\Q$ such that $\sL(q)=\sL(q')$ for $q\equiv q'$;
the abstract Kripke structure for $\cK$ and $\equiv$ is defined as the
Kripke structure $\cK_\cA=(\Q_\cA,q_\cA,\rightarrow_\cA,\sL_\cA)$
where
\begin{itemize}
\item $\Q_\cA=\{[q]_\equiv \mbar q\in \Q\}$ is the set of equivalence
  classes under $\equiv$,
\item $q_\cA=[q_\I]_\equiv$,
\item $[q]_\equiv \rightarrow_\cA [q']_\equiv$ if there is some
  $q_1\in [q]_\equiv$ and $q_1'\in[q']_\equiv$ such that
  $q_1\rightarrow q_1'$, and
\item $\sL_\cA([q]_\equiv) = \sL(q).$
\end{itemize}

\begin{figure}[t]
\begin{minipage}[b]{0.5\textwidth}
\begin{center}
\begin{pgfpicture}{0cm}{0cm}{5cm}{5cm}
\footnotesize
\pgfnodecircle{q0}[stroke]{\pgfxy(3.25,4.75)}{0.25cm}
\pgfnodecircle{q1}[stroke]{\pgfxy(1.75,4.75)}{0.25cm}
\pgfnodecircle{q2}[stroke]{\pgfxy(3.25,3.25)}{0.25cm}
\pgfnodecircle{q3}[stroke]{\pgfxy(4.75,4.75)}{0.25cm}
\pgfnodecircle{q4}[stroke]{\pgfxy(4.75,3.25)}{0.25cm}
\pgfnodecircle{q5}[stroke]{\pgfxy(1.75,3.25)}{0.25cm}
\pgfnodecircle{q6}[stroke]{\pgfxy(0.25,3.25)}{0.25cm}
\pgfnodecircle{q7}[stroke]{\pgfxy(0.25,1.75)}{0.25cm}
\pgfnodecircle{q8}[stroke]{\pgfxy(1.75,1.75)}{0.25cm}
\pgfnodecircle{q9}[stroke]{\pgfxy(3.25,1.75)}{0.25cm}
\pgfnodecircle{q10}[stroke]{\pgfxy(4.75,1.75)}{0.25cm}
\pgfnodecircle{q11}[stroke]{\pgfxy(3.25,0.25)}{0.25cm}
\pgfputat{\pgfnodecenter{q0}}{\pgfbox[center,center]{$q_0$}}
\pgfputat{\pgfnodecenter{q1}}{\pgfbox[center,center]{$q_1$}}
\pgfputat{\pgfnodecenter{q2}}{\pgfbox[center,center]{$q_2$}}
\pgfputat{\pgfnodecenter{q3}}{\pgfbox[center,center]{$q_3$}}
\pgfputat{\pgfnodecenter{q4}}{\pgfbox[center,center]{$q_4$}}
\pgfputat{\pgfnodecenter{q5}}{\pgfbox[center,center]{$q_5$}}
\pgfputat{\pgfnodecenter{q6}}{\pgfbox[center,center]{$q_6$}}
\pgfputat{\pgfnodecenter{q7}}{\pgfbox[center,center]{$q_7$}}
\pgfputat{\pgfnodecenter{q8}}{\pgfbox[center,center]{$q_8$}}
\pgfputat{\pgfnodecenter{q9}}{\pgfbox[center,center]{$q_9$}}
\pgfputat{\pgfnodecenter{q10}}{\pgfbox[center,center]{$q_{10}$}}
\pgfputat{\pgfnodecenter{q11}}{\pgfbox[center,center]{$q_{11}$}}
\pgfsetendarrow{\pgfarrowsingle}
\pgfnodeconnline{q0}{q3}
\pgfnodeconnline{q0}{q2}
\pgfnodeconnline{q0}{q5}
\pgfnodeconnline{q3}{q4}
\pgfnodeconnline{q2}{q9}
\pgfnodeconnline{q4}{q10}
\pgfnodeconnline{q5}{q8}
\pgfnodeconnline{q6}{q7}
\pgfnodeconnline{q7}{q11}
\pgfnodeconnline{q9}{q11}
\pgfnodeconnline{q10}{q11}
\pgfxycurve(4.56,4.875)(4,5.5)(2.5,5.5)(1.96,4.875)
\pgfxycurve(3,0.25)(0.25,0.25)(-1,1.75)(0.1,3.05)
\end{pgfpicture}
\end{center}
\caption{Kripke structure $\cKex$}
\label{fig:example-kripke}
\end{minipage}
\begin{minipage}[b]{0.5\textwidth}
\begin{center}
\begin{pgfpicture}{-1cm}{0cm}{3.5cm}{5cm}
\footnotesize
\pgfnodebox{q013}[stroke]{\pgfxy(1.75,4.75)}{$\{q_0,q_1,q_3\}$}{3pt}{3pt}
\pgfnodebox{q56}[stroke]{\pgfxy(0.25,3.25)}{$\{q_5,q_6\}$}{3pt}{3pt}
\pgfnodebox{q2}[stroke]{\pgfxy(1.75,3.25)}{$\{q_2\}$}{3pt}{3pt}
\pgfnodebox{q4}[stroke]{\pgfxy(3.25,3.25)}{$\{q_4\}$}{3pt}{3pt}
\pgfnodebox{q78}[stroke]{\pgfxy(0.25,1.75)}{$\{q_7,q_8\}$}{3pt}{3pt}
\pgfnodebox{q9}[stroke]{\pgfxy(1.75,1.75)}{$\{q_9\}$}{3pt}{3pt}
\pgfnodebox{q10}[stroke]{\pgfxy(3.25,1.75)}{$\{q_{10}\}$}{3pt}{3pt}
\pgfnodebox{q11}[stroke]{\pgfxy(1.75,0.25)}{$\{q_{11}\}$}{3pt}{3pt}
\pgfsetendarrow{\pgfarrowsingle}
\pgfnodeconnline{q013}{q56}
\pgfnodeconnline{q013}{q2}
\pgfnodeconnline{q013}{q4}
\pgfnodeconnline{q56}{q78}
\pgfnodeconnline{q2}{q9}
\pgfnodeconnline{q4}{q10}
\pgfnodeconnline{q78}{q11}
\pgfnodeconnline{q9}{q11}
\pgfnodeconnline{q10}{q11}
\pgfxycurve(2.55,4.75)(3,4.75)(2.7,5.5)(2.3,5)
\pgfxycurve(1.25,0.25)(-0.25,0.25)(-1,1.75)(-0.1,3)
\end{pgfpicture}
\end{center}
\caption{Its abstraction $\cKab$}
\label{fig:example-abstract}
\end{minipage}
\end{figure}

\begin{example}\rm
\label{exam:Kripke}
Consider the Kripke $\cKex$ structure given in Figure
\ref{fig:example-kripke} where $q_0$ is the initial state and the
state $q_{11}$ is labeled by proposition $P$ (no other state is
labeled by any proposition). Consider the equivalence relation
$\equiv$ which partitions the set
$\{q_0,q_1,q_2,q_3,q_4,q_5,q_6,q_7,q_8,q_9,q_{10},q_{11}\}$ into the
equivalence classes $\{q_0,q_1,q_3\}$, $\{q_2\}$, $\{q_4\}$,
$\{q_5,q_6\}$, $\{q_7,q_8\}$, $\{q_9\}$, $\{q_{10}\}$ and $q_{11}$.
Then the abstract Kripke structure, $\cKab$ for $\cKex$ and $\equiv$
is given by the Kripke structure in Figure
\ref{fig:example-abstract}. Here $\{q_0,q_1,q_3\}$ is the initial
state and $\{q_{11}\}$ is labeled by proposition $P$.

\end{example}

This construction is generalized for MDP's in
\cite{jl91,hut05,Dajjl01}.  To describe this generalized construction
formally, we first need to lift distributions on a set with an
equivalence relation $\equiv$ to a distribution on the equivalence
classes of $\equiv$~\footnote{It is possible to avoid lifting
  distributions if one assumes that each transition in the systems is
  uniquely labeled, and has the property that the target
  sub-probability measure has non-zero measure for at most one state.
  This does not affect the expressive power of the model and is used
  in~\cite{holmanns}. However the disadvantage is that the abstract
  model may be larger as fewer transitions will be collapsed.}.
\begin{definition}
  Given $\mu \in \subdistr{\Q}$ and an equivalence $\equiv$ on $\Q$,
  the lifting of $\mu$ (denoted by $[\mu]_\equiv$) to the set of
  equivalence classes of $\Q$ under $\equiv$ is defined as
  $[\mu]_\equiv ([q]_\equiv) = \mu(\{q'\in\Q \st q'\equiv q\})$.
\end{definition}
For an MDP $\cM=(\Q,q_\I, \delta, \sL)$, we will say a binary relation
$\equiv$ is an \emph{equivalence relation compatible with $\cM$}, if
$\equiv$ is an equivalence relation on $\Q$ such that $\sL(q)=\sL(q')$
for all $q\equiv q'$.  The abstract models used in our framework are
then formally defined as follows.
\begin{definition}
  Given a set of propositions $\AP$, let $\cM=(\Q,q_\I, \delta, \sL)$
  be an $\AP$ labeled MDP.  Let $\equiv$ be an equivalence relation
  compatible with $\cM$.  The {\it abstract MDP for $\cM$ with respect
    to the equivalence relation $\equiv$} is a MDP
  $\absmdp{\cM}{\equiv} = (\amdpcomp{\Q}{\equiv},
  \amdpcomp{q}{\equiv},\amdpcomp{\delta}{\equiv},\amdpcomp{\sL}{\equiv})$
  where
\begin{enumerate}
\item $\amdpcomp{\Q}{\equiv}=\{[q]_\equiv \mbar q\in \Q\}$ .
\item $\amdpcomp{q}{\equiv}=[q_\I]_\equiv$. 
\item $\amdpcomp{\delta}{\equiv}([q]_\equiv)=\{ \mu \st \exists q'\in
  [q]_\equiv \textrm{ and } \mu_1 \in \delta(q') \textrm{ such that }
  \mu=[\mu_1]_\equiv \}$.
\item $\amdpcomp{\sL}{\equiv}([q]_\equiv)=\sL(q)$.
\end{enumerate}
The elements of $\Q$ shall henceforth be called {\it concrete states}
and the elements $\amdpcomp{\Q}{\equiv}$ shall henceforth be called
{\it abstract states}.  The relation $\alrel{\equiv} \subseteq \Q
\times \amdpcomp{\Q}{\equiv}$ defined as
$\alrel{\equiv}=\{(q,[q]_\equiv)\st q\in \Q\}$ shall henceforth be
called the {\it abstraction relation}. The relation
$\gmrel{\equiv}\subseteq \Q_\cA\times \Q $ defined as
$\gmrel{\equiv}=\{([q]_\equiv,q')\st [q]_\equiv \in
\amdpcomp{\Q}{\equiv}, q\equiv q'\}$ shall henceforth be called {\it
  concretization relation}.
\end{definition}

\begin{remark}
  The relation $\alrel{\equiv}$ is total and functional and hence
  represents a function $\alpha$ which is often called the {\it
    abstraction map} in literature. Please note that one can define
  the equivalence $\equiv$ via the function $\alpha$.  The relation
  $\gmrel{\equiv}$ is total (not necessarily functional) and hence
  represents a map into the power-set $2^\Q$. The function
  $\gamma:\amdpcomp{\Q}{\equiv}\rightarrow 2^\Q$ defined as
  $\gamma(a)=\gmrel{\equiv}(a)$ is often called the {\it
    concretization map} in literature.
\end{remark}

We conclude this section by making a couple of observations about the
construction of the abstract MDP. First notice that the abstract MDP
$\absmdp{\cM}{\equiv}$ has been defined to ensure that it simulates
$\cM$ via the canonical simulation relation $\alrel{\equiv}$.  Next,
we show that we can obtain a ``refinement'' of the abstract MDP
$\absmdp{\cM}{\equiv}$, by considering the abstraction of $\cM$ with
respect to another equivalence $\simeq$ that is finer than $\equiv$.
This is stated next.

\begin{definition}
  Let $\cM=(\Q,q_\I, \delta, \sL)$ be an MDP over the set of atomic
  propositions $\AP$. Further let $\equiv$ and $\simeq$ be two
  equivalence relations compatible with $\cM$ such that $\simeq
  \subseteq \equiv$. The abstract MDP $\absmdp{\cM}{\simeq}$ is said
  to be a {\it refinement} of $\absmdp{\cM}{\equiv}$. The relation
  $\alrel{\simeq,\equiv}\subseteq \amdpcomp{\Q}{\simeq} \times
  \amdpcomp{\Q}{\equiv}$ defined as $\{([q']_\simeq,[q]_\equiv)\st
  [q']_{\simeq}\subseteq[q]_\equiv\}$ is said to be a refinement
  relation for $(\absmdp{\cM}{\simeq},\absmdp{\cM}{\equiv})$.
\end{definition}
The following is an immediate consequence of the definition.
\begin{proposition}
  Let $\simeq$ and $\equiv$ be two equivalence relations compatible
  with the MDP $\cM$ such that $\simeq \subseteq \equiv$. Recall that
  the refinement relation for
  $(\absmdp{\cM}{\simeq},\absmdp{\cM}{\equiv})$ is denoted by
  $\alrel{\simeq,\equiv}$. Then $\alrel{\simeq,\equiv}$ is a canonical
  simulation and $\alrel{\simeq,\equiv}\circ
  \alrel{\simeq}=\alrel{\equiv}$.
\end{proposition}

\section{Counterexample Guided Refinement}
\label{sec:refinement}

As described in Section \ref{sec:abs}, in our framework, an MDP $\cM$
will be abstracted by another MDP $\absmdp{\cM}{\equiv}$ defined on
the basis of an equivalence relation $\equiv$ on the states of $\cM$.
Model checking $\absmdp{\cM}{\equiv}$ against a safety property
$\psi_S$ will either tell us that $\psi_S$ is satisfied by
$\absmdp{\cM}{\equiv}$ (in which case, it is also satisfied by $\cM$
as shown in Lemma~\ref{thm:lem-pctl-safety}) or it is not. If
$\absmdp{\cM}{\equiv}\not\sat\psi_S$ then $\absmdp{\cM}{\equiv}$ can
be analyzed to obtain a minimal counterexample $(\cE,\rel\inj)$, using
the algorithm in Theorem~\ref{thm:min}. The counterexample
$(\cE,\rel\inj)$ must be analyzed to decide whether $(\cE,\rel\inj)$
proves that $\cM$ fails to satisfy $\psi_S$, or the counterexample is
\emph{spurious} and the abstraction (or rather the equivalence
relation $\equiv$) must be refined to ``eliminate'' it. In order to
carry out these steps, we need to first identify what it means for a
counterexample to be \emph{valid and consistent} for $\cM$, describe
and analyze an algorithm to check validity, and then demonstrate how
the abstraction can be refined if the counterexample is spurious. In
this section, we will outline our proposal to carry out these steps.
We will frequently recall how these steps are carried out in the
non-probabilistic case through a running example to convince the
reader that our definitions are a natural generalization to the
probabilistic case.

\subsection{Checking Counterexamples}

Checking if a counterexample proves that the system $\cM$ fails to
meet its requirements $\psi_S$, intuitively, requires one to check if
the ``behavior'' (or behaviors) captured by the counterexample are
indeed exhibited by the system. The formal concept that expresses when
a systems exhibits certain behaviors is \emph{simulation}. Thus, one
could potentially consider defining a valid counterexample to be one
that is simulated by the MDP $\cM$.  However, as we illustrate in this
section, the notion of valid counterexamples that is used in the
context of non-probabilistic systems~\cite{cgjlv00,cjlv02} is
stronger. We, therefore, begin by motivating and formally defining
when a counterexample is valid and consistent
(Section~\ref{sec:valid-const}), and then present and analyze the
algorithm for checking validity (Section~\ref{sec:gen-check-valid}).

\subsubsection{Validity and Consistency of Counterexamples}
\label{sec:valid-const}

In the context of non-probabilistic systems, a valid counterexample is
not simply one that is simulated by the original system. This is
illustrated by the following example; we use this to motivate our
generalization to probabilistic systems.

\begin{figure}[t]
\begin{center}
\begin{pgfpicture}{-1cm}{0cm}{3.5cm}{5cm}
\footnotesize
\pgfnodebox{q013}[stroke]{\pgfxy(1.75,4.75)}{$\{q_0,q_1,q_3\}$}{3pt}{3pt}
\pgfnodebox{q56}[stroke]{\pgfxy(0.25,3.25)}{$\{q_5,q_6\}$}{3pt}{3pt}
\pgfnodebox{q2}[stroke]{\pgfxy(1.75,3.25)}{$\{q_2\}$}{3pt}{3pt}
\pgfnodebox{q4}[stroke]{\pgfxy(3.25,3.25)}{$\{q_4\}$}{3pt}{3pt}
\pgfnodebox{q78}[stroke]{\pgfxy(0.25,1.75)}{$\{q_7,q_8\}$}{3pt}{3pt}
\pgfnodebox{q9}[stroke]{\pgfxy(1.75,1.75)}{$\{q_9\}$}{3pt}{3pt}
\pgfnodebox{q10}[stroke]{\pgfxy(3.25,1.75)}{$\{q_{10}\}$}{3pt}{3pt}
\pgfnodebox{q11}[stroke]{\pgfxy(1.75,0.25)}{$\{q_{11}\}$}{3pt}{3pt}
\pgfsetendarrow{\pgfarrowsingle}
\pgfxycurve(2.55,4.75)(3,4.75)(2.7,5.5)(2.3,5)
\pgfxycurve(1.25,0.25)(-0.25,0.25)(-1,1.75)(-0.1,3)
\pgfsetlinewidth{1pt}
\pgfsetdash{{0.1cm}{0.1cm}}{0cm}
\pgfnodeconnline{q013}{q2}
\pgfnodeconnline{q2}{q9}
\pgfnodeconnline{q9}{q11}
\pgfsetdash{{0.5cm}{0.1cm}}{0cm}
\pgfnodeconnline{q013}{q4}
\pgfnodeconnline{q4}{q10}
\pgfnodeconnline{q10}{q11}
\pgfsetdash{{0.1cm}{0.1cm}{0.1cm}{0.1cm}{0.5cm}{0.1cm}}{0cm}
\pgfnodeconnline{q013}{q56}
\pgfnodeconnline{q56}{q78}
\pgfnodeconnline{q78}{q11}
\end{pgfpicture}
\end{center}
\caption{The three counterexamples $\cKcexa$ (shown with short dashed edges), 
$\cKcexb$ (shown with long dashed edges), and $\cKcexc$ 
(shown with short and long dashed edges)}
\label{fig:counterex}
\end{figure}

\begin{example}\rm
\label{exam:check} 
Recall the Kripke structure $\cKex$ given in Example \ref{exam:Kripke}
along with the abstraction $\cKab$ (these structures are given in
Figures~\ref{fig:example-kripke} and~\ref{fig:example-abstract},
respectively).  The $\mathsf{LTL}$ safety-property $\Box (\neg P)$ is
violated by $\cKab$. For such safety properties, counterexamples are
just paths in $\cKab$ (which of course can be viewed as Kripke
structures in their own right). The counterexample generation
algorithms in \cite{cgjlv00,cjlv02} could possibly generate any one of
three paths in $\cK_{ab}$ shown in Fig~\ref{fig:counterex}:
$\cKcexa=\{q_0,q_1,q_3\}\rightarrow\{q_2\}\rightarrow \{q_9\}
\rightarrow \{q_{11}\}$,
$\cKcexb=\{q_0,q_1,q_3\}\rightarrow\{q_4\}\rightarrow \{q_{10}\}
\rightarrow \{q_{11}\}$ and
$\cKcexc=\{q_0,q_1,q_3\}\rightarrow\{q_5,q_6\}\rightarrow\{q_7,q_8\}
\rightarrow \{q_{11}\}.$ Now each of the counter-examples
$\cKcexa,\cKcexb$ and $\cKcexc$ is simulated by $\cKex$ because
$\cKex$ has a path, starting from the initial state, having 4 states,
where only the fourth state satisfies proposition $P$. However, the
algorithm outlined in~\cite{cgjlv00,cjlv02} only considers $\cKcexa$
to be valid. In order to see this, let us recall how the algorithm
proceeds. The algorithm starts from the last state of the
counterexample and proceeds backwards, checking at each point whether
any of the concrete states corresponding to the abstract state in the
counterexample can exhibit the counterexample from that point
onwards. Thus, $\cKcexb$ is invalid because $q_0$ does not have a
transition to $q_4$ and $\cKcexc$ is invalid because none of
$q_0,q_1$, or $q_3$ have a transition to $q_6$ (the only state among
$q_5$ and $q_6$ that can exhibit $\{q_5,q_6\}\rightarrow\{q_7,q_8\}
\rightarrow \{q_{11}\}$).
\end{example}

The example above illustrates that to check validity, the algorithm
searches for a simulation relation, wherein each (abstract) state of
the counterexample is mapped to one of the concrete states that
correspond to it, rather than an arbitrary simulation relation. Thus
the ``proof'' for the validity of a counterexample in a concrete
system, must be ``contained'' in the proof that demonstrates the
validity of the counterexample in the abstract system. Based on this
intuition we formalize the notion of when a counterexample is valid
and consistent.

\begin{definition}
  Let $\cM$ be an MDP with set of states $\Q$, and $\equiv$ be an
  equivalence relation that is compatible with $\cM$. Let $\psi_S$ be
  a {\PCTL}-safety formula such that
  $\absmdp{\cM}{\equiv}\not\sat\psi_S$ and let $(\cE,\cR_0)$ be a
  counterexample for $\absmdp{\cM}{\equiv}$ and $\psi_S$ with set of
  states $\Q_\cE$.  We say that the counterexample $(\cE,\cR_0)$ is
  {\it valid} and {\it consistent with $(\cM,\equiv)$} if there is a
  relation $\cR\subseteq \Q_\cE\times \Q$ such that
  \begin{enumerate}
  \item $\cR$ is a canonical simulation (validity); and
  \item $\alrel{\equiv}\circ \cR  \subseteq \cR_0$ (consistency).
  \end{enumerate}
  The relation $\cR$ is said to be a {\it validating simulation}.  If
  no such $\cR$ exists then $(\cE,\cR_0)$ is said to be {\it invalid
    for $(\cM,\equiv)$}.
\end{definition}
The above definition provides one technical reason for why it is
convenient to view a counterexample as not just an MDP but rather as
an MDP along with a simulation relation; we will see another
justification for this when we discuss refinement.

\begin{remark}
  When the counterexample $(\cE,\rel\inj)$ is generated as in
  Theorem~\ref{thm:min}, $\Q_\cE\subseteq \{\bar{a}\st a \in
  \amdpcomp{\Q}{\equiv}\}$ and the relation $\rel\inj=\{(\bar{a},a)\st
  \bar{a} \in \Q_\cE \}$. In this case, please note that consistency
  is equivalent to requiring that $\cR\subseteq \{(\bar{a},q)\st q\in
  \gmrel{\equiv}(a)\}.$ In other words, consistency is equivalent to
  requiring that $\cR\subseteq \gmrel{\equiv}\circ \rel\inj.$
\end{remark}

We conclude this section by showing that for minimal counterexamples
$(\cE,\cR_0)$ the containment in the consistency requirement can be
taken to be equality.
\begin{proposition}
\label{prop:minimal-eq}
  Let $\cM$ be a MDP, $\equiv$ an equivalence relation compatible with
  $\cM$ and $\psi_S$ be a safety formula such that
  $\absmdp{\cM}{\equiv}\not\sat\psi_S$.  If $(\cE,\cR_0)$ is a minimal
  counterexample for $\absmdp{\cM}{\equiv}$ and $\psi_S$ and
  $(\cE,\cR_0)$ is consistent and valid for $(\cM,\equiv)$ with
  validating simulation $\cR$ then $ \alrel{\equiv}\circ \cR = \cR_0.$
\end{proposition}
\begin{proof}
  We have that $\cR$ is a simulation and $\alrel{\equiv}\circ
  \cR\subseteq\cR_0$.  Since $\alrel{\equiv}$ and $\cR$ are
  simulations, so is $\alrel{\equiv}\circ\cR$. Also since $\cE\not\sat
  \psi_S$, we get that $(\cE,\alrel{\equiv}\circ \cR)$ is also a
  counterexample for $\absmdp{\cM}{\equiv}$ and $\psi_S$.  Thus, if
  $\alrel{\equiv}\circ \cR\subsetneq \cR_0$, then
  $(\cE,\alrel{\equiv}\circ \cR)$ is not minimal. Hence
  $\alrel{\equiv}\circ \cR = \cR_0.$
\end{proof}

\subsubsection{Algorithm to check Validity of Counterexamples}
\label{sec:gen-check-valid}

We now present the algorithm to check the validity and consistency of
a counterexample. We will assume that the counterexample is a minimal
one, generated by the algorithm in Theorem~\ref{thm:min}. Thus the
counterexample is of the form $(\cE,\rel\inj)$, where the set of
states $\Q_\cE$ is a subset of $\{\bar{a}\st a \in
\amdpcomp{\Q}{\equiv}\}$ and the relation $\rel\inj$ is
$\{(\bar{a},a)\st \bar{a}\in \Q_\cE\}$.  The algorithm for
counterexample checking is then the standard simulation checking
algorithm~\cite{bem00} that computes the validating simulation through
progressive refinement, except that in our case we start with $R =
\{(\bar{a},q)\; |\; q \in \rel\equiv^\gamma(a)\}$.  The algorithm is shown in
Figure~\ref{fig:validity}. Please note that for the rest of the paper
(and in the algorithm) by $\mu\simd{R_{old}}\mu'$ we mean
$\mu\simd{\id_{\Q}\cup R_{old}\cup \id_{\Q_\cE}}\mu'.$

\begin{figure}[t]
\begin{center}
\begin{tabular}{|l|}
\hline
\\
Initially $R = \{(\bar{a},q)\; |\; q \in \gmrel{\equiv}(\bar{a})\}$ and $R_{old} = \emptyset$\\
while $(R_{old} \neq R)$ do \\
\hspace*{0.5cm}$R_{old} = R$\\
\hspace*{0.5cm}For each state $\bar{a} \in \Q_\cE$ and each $\mu \in \delta_\cE(\bar{a})$ do\\
            
\hspace*{1cm}$R = \{(\bar{b},q)\in R\,|\, b\ne a\} \cup \{ (\bar{a},q)\in R\,|\,
\exists \mu' \in \delta(q).\ \mu
\simd{R_{old}} \mu'\}$\\
\hspace*{1cm}If $\cR(\bar{a})=\emptyset$ 
then return (``invalid'',$\bar{a}$, $\mu$, $R_{old}$, $R$)\\
\hspace*{1cm}If $\bar{a}=q_\cE$ and $q_\I\notin R(\bar{a}) $ then return (``invalid'',$\bar{a}$, $\mu$, $R_{old}$, $R$)\\
\hspace*{0.5cm} od $\leftarrow$ end of For loop\\
od $\leftarrow$ end of while loop\\
Return (``valid'')\\
\hline
\end{tabular}
\end{center}
\caption{Algorithm for checking validity and consistency of counterexamples}
\label{fig:validity}
\end{figure}

We will now show that the algorithm in Figure \ref{fig:validity} is
correct. We start by showing that the algorithm terminates.
\begin{proposition}
\label{prop:terminate1}
The counterexample checking algorithm shown in Figure
\ref{fig:validity} terminates.
\end{proposition}
\begin{proof}
  Let $\cR^n_0$ and $\cR^n_1$ be respectively the relations denoted by
  the variable $R$ at the beginning and the end of the $n$-th
  iteration of the while loop. A simple inspection of the algorithm
  tells us that either $\cR^n_1=\cR^n_0$ or $\cR^n_1\subsetneq
  \cR^n_0$. If $\cR^n_1=\cR^n_0$ then the while loop terminates (and
  returns ``valid'').  Otherwise the size of relation denoted by
  variable $R$ decreases by at least one.  Hence, if $\cR^n_1$ is
  never equal to $\cR^n_0$ then it must be case that
  $\cR^n_1(\bar{a})$ becomes empty for some $n$ and some $\bar{a}\in
  \Q_\cE$ and then the algorithm terminates.
\end{proof}
 
We now show that if the algorithm returns ``valid'' then the
counterexample $(\cE,\rel\inj)$ is valid and consistent.
\begin{proposition}
\label{prop:validity1}
If the algorithm in Figure~\ref{fig:validity} returns ``valid'' then
the counterexample $(\cE,\rel\inj)$ is valid and consistent for
$(\cM,\equiv).$
\end{proposition}
\begin{proof}
  Please note that the algorithm returns ``valid'' only when the while
  loop terminates. At that point the variables $R_{old}$ and $R$
  denote the same relation, which we shall call $\cR$ for the rest of
  the proof.  Please note as $\cR\subseteq \{(\bar{a},q)\; |\; q \in
  \gmrel{\equiv}(\bar{a})\}$, $\alrel{\equiv}\circ \cR \subseteq \rel\inj.$
  Hence, the result will follow if we can show that $\cR$ is a
  canonical simulation. Consider the last iteration of the while loop.
  Now, a simple inspection of the algorithm says that the variable $R$
  does not change its value in this iteration.  The only way this is
  possible is if for each $\bar{a}\in \Q_\cE$, $\mu \in
  \delta_\cE(\bar{a})$ and each $(\bar{a},q) \in \cR$, there exists
  $\mu'\in \delta(q)$ such that $\mu \simd{\cR} \mu'.$ Also, $q_\I \in \cR(q_\cE).$
   Thus $\cR$ is a
  canonical simulation.
\end{proof}

We now show that the if $(\Q_\cE,\rel\inj)$ is valid and consistent,
then the algorithm in Figure \ref{fig:validity} must return ``valid''.
\begin{proposition}
\label{prop:validity2}
If the counterexample $(\Q_\cE,\rel\inj)$ is valid and consistent with
$(\cM,\equiv)$ then the algorithm in Figure \ref{fig:validity} returns
``valid''.
\end{proposition}
\begin{proof}
  Assume that the counterexample $(\Q_\cE,\rel\inj)$ is valid and
  consistent. Thus there is a canonical simulation $\cR\subseteq
  \Q_\cE\times \amdpcomp{\Q}{\equiv}$ such that $\alrel{\equiv}\circ
  \cR=\rel\inj$ (equality is a consequence of minimality; see
  Proposition~\ref{prop:minimal-eq}). We make the following
  observations:
\begin{enumerate}
\item As $\alrel{\equiv}\circ \cR=\rel\inj$, $\cR\subseteq
  \{(\bar{a},q)\st q\in \gmrel{\equiv}(a)\}.$
\item For each $\bar{a} \in \Q_\cE$, $\cR(\bar{a})\ne \emptyset$
  (otherwise $(\bar{a},a)$ will be present in $\rel\inj$ but not in
  $\alrel{\equiv}\circ \cR$).
\item $q_\cE \cR q_\I$ ($\cR$ is a simulation).
\item For each $\bar{a}\in \Q_\cE$, each $\mu \in\delta(\bar{a})$ and
  each $\bar{a}\R q$ there exists a $\mu'\in \delta(q)$ such that
  $\mu\simd{\cR}\mu'$ ($\cR$ is a simulation).
\item For any relation $\cR_1\subset \{(\bar{a},q)\st q\in
  \rel\gamma(a)\} $ such that $\cR\subset \cR_1$, $\mu_a
  \in\delta(\bar{a})$, $\mu_q\in \delta(q)$ we have that
  $\mu_a\simd{\cR}\mu_q$ implies that $\mu_a\simd{\cR_1}\mu_q.$
\end{enumerate}
Now, the first observation above implies that in the algorithm in
Figure \ref{fig:validity} initially $\cR$ is contained within the
relation denoted by the variable $R$. $\cR$ is also contained in the
relation denoted by the variable $R_{old}$, the first time the
variable $R_{old}$ takes  a non-empty value. From this point on, we
claim $\cR$ is always contained in the relations $R_{old}$ and $R$.
This claim is a consequence of the fourth and the fifth observations
which ensure that every time $R$ is updated, $\cR$ is contained in the
relation denoted by $R$. Finally note that second and third
observations ensure that the algorithm can never declare the
counterexample to be invalid and hence by termination
(Proposition~\ref{prop:terminate1}), the algorithm must return
``valid''.
\end{proof}

A careful analysis of the special structure of the validating
simulation yields better bounds than that reported in~\cite{bem00} for
general simulation.

\begin{theorem}
  Let $\cM$ be an MDP, $\equiv$ an equivalence relation compatible
  with $\cM$, and $\psi_S$ a safety property such that
  $\absmdp{\cM}{\equiv}\not\sat\psi_S$. Let $(\cE,\rel\inj)$ be a
  counterexample for $\absmdp{\cM}{\equiv}$ and $\psi_S$ generated
  using Theorem~\ref{thm:min}. Then the algorithm in
  Figure~\ref{fig:validity} returns ``valid'' iff $(\cE,\rel\inj)$ is
  valid and consistent with $(\cM,\equiv)$.  Let $n_\cM$ and $m_\cM$
  be the number of vertices and edges, respectively, in the underlying
  labeled graph $\lungrp{\cM}$. The algorithm shown in
  Figure~\ref{fig:validity} runs in time $O(n_\cM^2 m_\cM^2)$.
\end{theorem}
\begin{proof} 
  Thanks to Propositions \ref{prop:validity1} and
  \ref{prop:validity2}, the result will follow if we show that the
  running time of the algorithm is $O(n_\cM^2 m_\cM^2)$.
  
  Observe that the outermost while loop runs for as long as $R$
  changes.  If for each $\bar{a}\in \Q_\cE$, we define $s_a =
  |\gmrel{\equiv}(a)|=|\gmrel{\equiv}\circ \rel\inj(\bar{a})|$, a
  bound on the number of iterations of the outermost loop is
  $\sum_{\bar{a} \in \Q^\cE} s_a = n_\cM$, as each state of $\Q$
  belongs to at most one $\gmrel{\equiv}(a)$. Next let us define $d_a$
  to be number of outgoing edges from the set $\gmrel{\equiv}(a)$ and
  $d_q$ to be the out-degree of $q$ in $\lungrp{\cM}$. Clearly for
  each state $\bar{a}\in \Q_\cE$, the total number of tests of the
  form $\mu \simd{R_{old}} \mu'$ is bounded by $\sum_{q \in
    \gmrel{\equiv}(a)}d_ad_q = d_a^2$. Thus, the total number of tests
  $\mu \simd{R_{old}} \mu'$ in a single iteration of the outermost
  loop is bounded by $\sum_{\bar{a} \in \Q_\cE}d_a^2 \leq m_\cM^2$.
  Now, because each state of $\Q$ belongs to at most one
  $\gmrel{\equiv}\circ \rel\inj(\bar{a})$, the test $\mu
  \simd{R_{old}} \mu'$ simply requires one to check that for each
  $\bar{b}\in \Q_\cE$, $\mu(\bar{b}) \leq \sum_{q: (\bar{a},q) \in
    R_{old}} \mu'(q)$ and thus can be done in $O(n_\cM)$ time. Thus we
  don't need to compute flows in bipartite networks, as is required in
  the general case~\cite{bem00}.  The total running time is,
  therefore, $O(n_\cM^2m_\cM^2)$.
\end{proof}
We make some observations about the validity checking algorithm.
\begin{enumerate}
\item For linear time properties and non-probabilistic systems,
  checking validity of a counterexample simply determines if the first
  state in the counterexample can be simulated by the initial state of
  the system by going backwards from the last state of the
  counterexample. The same idea can be exploited for probabilistic
  systems as well if the underlying unlabeled graph of the
  counterexample $\cE$ is a tree (or more generally a DAG); the
  resulting algorithm will depend on the height of the counterexample
  and cut the running time of the algorithm by a factor of $n_\cM$.
\item Theorem~\ref{thm:tree-cntr-strict-liveness} observes that for
  weak safety formulas, counterexamples whose underlying graph is a
  tree can be found by ``unrolling'' the minimal MDP counterexample.
  Instead of first explicitly unrolling the counterexample and then
  checking, one can unroll the counterexample ``on the fly'' while
  checking validity. This algorithm is presented in
  Section~\ref{sec:onthefly}, after our discussion on refinement so as
  to not interrupt the flow. The crucial idea is to decide when to
  stop unrolling which is made by keeping track of the satisfaction of
  subformulas at various states. The running time of the algorithm
  will be $O(h\cdot n_\cM m_\cM^2)$, where $h$ is the height of the
  unrolled tree. Thus depending on $n_\cM$ and $h$, one could either
  compute the actual simulation relation, or simply check whether the
  tree of height $h$ is simulated.
\item One can construct the graph of maximal strongly connected
  components of $\lungrp{\cE}$, and compute the simulation relation on
  each maximal strongly connect component, in the order of their
  topological sort. While this new algorithm will not yield better
  asymptotic bounds, it may work better in practice.
\end{enumerate}

To complete the description of the CEGAR approach, all we need to do
is describe the refinement step. However, before we proceed, we
describe a result and give some notations for the case when the
counterexample generated by the algorithm in Figure~\ref{fig:min} is
declared by the counterexample checking algorithm to be invalid.
\begin{proposition}
\label{prop:violation}
If the algorithm in Figure~\ref{fig:validity} returns (``invalid'',
$\bar{a}$, $\mu, R_{old}$, $R$), then for all $q \in
R_{old}(\bar{a})\setminus \R(\bar{a})$ and all $\mu_1\in \delta(q)$,
$\mu \not \simd{R_{old}} \mu_1.$
\end{proposition}
\begin{proof}
Immediate consequence of the algorithm.
\end{proof}

\begin{notation}
  If the algorithm in Figure~\ref{fig:validity} returns (``invalid'',
  $\bar{a}$, $\mu, R_{old}$, $R$) then
\begin{itemize}
\item $\bar{a}$ is said to be an {\it invalidating abstract state};
\item $\mu\in\delta_{\Q_\cE}(\bar{a})$ is said to be an {\it
    invalidating transition}; and
\item the pair $(R_{old},R)$ is said to be the {\it invalidating
    witness} for $\bar{a}$ and $\mu.$
\end{itemize}
\end{notation}

\subsection{Refining Abstractions}

The last step in the abstraction-refinement loop is to refine the
abstraction in the case when the counterexample is invalid. The
algorithm in Figure~\ref{fig:validity}, concludes the invalidity of
the counterexample, when it finds some abstract state $a$ such that
$\bar{a}$ is a state of the counterexample and $\bar{a}$ is not
simulated by any concrete state in $\rel\gamma(a)$, or when $\bar{a}$
is the initial state of the counterexample and it is not simulated by
the initial state of $\cM$. At this point, we will refine the
abstraction by refining the equivalence $\equiv$ that was used to
construct the abstract model in the first place. The goal of the
refinement step is for it to be ``counterexample guided''. The ideal
situation is one where the spurious counterexample is ``eliminated''
by the refinement step. However, as we remind the reader, this is not
achieved in the CEGAR approach for non-probabilistic systems. We,
therefore, begin (Section~\ref{sec:good-refine}) by motivating and
defining the notion of a ``good refinement''. We show that good
refinements do indeed lead to progress in the CEGAR approach. After
this, in Section~\ref{sec:refine-algo}, we present a refinement
algorithm along with a proof that it results in good refinements.

\subsubsection{Good Refinements}
\label{sec:good-refine}

We begin by recalling the refinement step in the CEGAR approach for
non-probabilistic systems through an example, to demonstrate that
refinement does not lead to the elimination of the counterexample. The
example, however, motivates what the refinement step does indeed
achieve, leading us to the notion of good refinements.

\begin{figure}[t]
\begin{center}
\begin{pgfpicture}{-2cm}{0cm}{3.5cm}{5cm}
\footnotesize
\pgfnodebox{q013}[stroke]{\pgfxy(1.75,4.75)}{$\{q_0,q_1,q_3\}$}{3pt}{3pt}
\pgfnodebox{q6}[stroke]{\pgfxy(-1.25,3.25)}{$\{q_6\}$}{3pt}{3pt}
\pgfnodebox{q5}[stroke]{\pgfxy(0.25,3.25)}{$\{q_5\}$}{3pt}{3pt}
\pgfnodebox{q2}[stroke]{\pgfxy(1.75,3.25)}{$\{q_2\}$}{3pt}{3pt}
\pgfnodebox{q4}[stroke]{\pgfxy(3.25,3.25)}{$\{q_4\}$}{3pt}{3pt}
\pgfnodebox{q78}[stroke]{\pgfxy(0.25,1.75)}{$\{q_7,q_8\}$}{3pt}{3pt}
\pgfnodebox{q9}[stroke]{\pgfxy(1.75,1.75)}{$\{q_9\}$}{3pt}{3pt}
\pgfnodebox{q10}[stroke]{\pgfxy(3.25,1.75)}{$\{q_{10}\}$}{3pt}{3pt}
\pgfnodebox{q11}[stroke]{\pgfxy(1.75,0.25)}{$\{q_{11}\}$}{3pt}{3pt}
\pgfsetendarrow{\pgfarrowsingle}
\pgfnodeconnline{q013}{q5}
\pgfnodeconnline{q013}{q2}
\pgfnodeconnline{q013}{q4}
\pgfnodeconnline{q5}{q78}
\pgfnodeconnline{q6}{q78}
\pgfnodeconnline{q2}{q9}
\pgfnodeconnline{q4}{q10}
\pgfnodeconnline{q78}{q11}
\pgfnodeconnline{q9}{q11}
\pgfnodeconnline{q10}{q11}
\pgfxycurve(2.55,4.75)(3,4.75)(2.7,5.5)(2.3,5)
\pgfxycurve(1.25,0.25)(-1.25,0.25)(-1.25,1.75)(-1.25,3)
\end{pgfpicture}
\end{center}
\caption{The abstraction $\cKab'$ }
\label{fig:ab2}
\end{figure}
\begin{example}\rm
\label{exam:refine}
As in Example \ref{exam:check}, consider the Kripke structure $\cKex$
from Example \ref{exam:Kripke} along with the abstraction $\cKab$
(these structures are also given in Figures~\ref{fig:example-kripke}
and~\ref{fig:example-abstract}). The $\mathsf{LTL}$ safety-property
$\phi=\Box (\neg P)$ is violated by $\cKab$ and the counterexample
generation algorithm may generate the spurious counterexample
$\cKcexc=\{q_0,q_1,q_3\}\rightarrow\{q_5,q_6\}
\rightarrow\{q_7,q_8\}\rightarrow \{q_{11}\}.$ Now, the counterexample
checking algorithm in \cite{cgjlv00,cjlv02} starts from the last state
of the counterexample and proceeds backwards, checking at each point
whether any of the concrete states corresponding to the abstract state
in the counterexample can exhibit the counterexample from that point
onwards. The algorithm finds that there is a path $q_6\rightarrow q_7
\rightarrow q_{11}$ in $\cKex$ which simulates
$\{q_5,q_6\}\rightarrow\{q_7,q_8\}\rightarrow \{q_{11}\}$ but finds
that there is no transition from $\{q_0,q_1,q_3\}$ to $q_6$. At this
point, it declares the counterexample to be invalid. The refinement
step then breaks the equivalence class
$\post(\{q_0,q_1,q_3\})=\{q_5,q_6\}$ into $\{q_6\}$ and $\{q_5\}.$ The
resulting abstraction $\cKab'$ shown in Figure \ref{fig:ab2} still has
the ``spurious'' counterexample
$\cKcexc'=\{q_0,q_1,q_3\}\rightarrow\{q_5\}\rightarrow\{q_7,q_8\}
\rightarrow\{q_{11}\}$ ``contained'' within $\cKcexc.$
\end{example}


Though, in Example~\ref{exam:refine}, the counterexample is not
eliminated by the refinement, progress is nonetheless made.
Considering the example carefully, one notes that breaking
$\{q_5,q_6\}$ could have yielded two possible new paths --
$\cKcexc'=\{q_0,q_1,q_3\}\rightarrow\{q_5\}\rightarrow
\{q_7,q_8\}\rightarrow\{q_{11}\}$ and
$\cKcexc''=\{q_0,q_1,q_3\}\rightarrow\{q_6\}\rightarrow
\{q_7,q_8\}\rightarrow\{q_{11}\}.$ However, only one path $\cKcexc'$
is a simulated by $\cKab'$ while the other path $\cKcexc''$ is not
simulated by $\cKab'$ and hence has been ``eliminated''. Thus, what is
eliminated is at least one simulation relation that is ``contained''
in the original spurious counterexample. We capture this concept for
MDPs as follows.

\begin{definition}
  Let $\cM$ be an MDP with states $\Q$, $\simeq$ and $\equiv$ be
  equivalence relations on $\Q$ compatible with $\cM$ such that
  $\simeq \subseteq \equiv$, and $\psi_S$ be a safety property. Let
  $(\cE,\cR)$ be a counterexample for $\absmdp{\cM}{\equiv}$ and
  $\psi_S$, where $\Q_\cE$ is the set of states of $\cE$ with initial
  state $q_\cE$. Finally let $\amdpcomp{q}{\simeq}$ by the initial
  state of $\absmdp{\cM}{\simeq}$. We say that $\simeq$ is a {\it good
    $\equiv$-refinement} for $(\cE,\cR)$ if there is \emph{some}
  $\cR'\subseteq \Q_\cE\times \amdpcomp{\Q}{\simeq}$ such that
  $(q_\cE, \amdpcomp{q}{\simeq})\in \cR'$, $\alrel{\simeq,\equiv}\circ
  \cR'= \cR$ but $\cR'$ is not a canonical simulation (of $\cE$ by
  $\absmdp{\cM}{\simeq}$).  If no such $\cR'$ exists, we say that
  $\simeq$ is a {\it bad $\equiv$-refinement} for $(\cE,\cR)$.
\end{definition}  
Intuitively, a good $\equiv$-refinement $\simeq$ ensures that
$(\cE,\cR')$ is not a counterexample for $\absmdp{\cM}{\simeq}$ and
$\psi_S$. Observe that the conditions on $\cR'$ ensure that $\cR'$ is
one of the possible proofs that $\cM$ violates $\psi_S$ ``contained''
within the counterexample $(\cE,\cR)$. This presents yet another
justification for formally treating the simulation relation (or proof)
as part of the notion of a counterexample. In the absence of the
simulation relation, it is difficult justify why refinement is
``counterexample guided'' given that the behavior (i.e., $\cE$) itself
may not be eliminated.

\begin{remark}
  Before presenting the consequences of good refinements, we would
  like examine the formal definition more carefully, in order to
  highlight the subtle reasons why all the points in the  definition are needed.
\begin{enumerate}
\item Observe that $\cR'$ satisfying $\alrel{\simeq,\equiv}\circ\cR' =
  \cR$ always exist: take $\cR'$ to be any relation such that for each
  $q_0\in \Q_\cE$, $\cR'(q_0)= \cup_{[q]_\equiv \in \cR(q_0)}
  X_{[q]_\equiv}$ where $X_{[q]_\equiv}$ is any non-empty subset of
  $\{[q_1]_{\simeq} \in \amdpcomp{\Q}{\simeq}\st [q_1]_{\simeq}
  \subseteq [q]_{\equiv} \}.$ Further, any $\cR'$ such that
  $\alrel{\simeq,\equiv}\circ \cR'= \cR$ must be of this form.
\item We demand $\alrel{\simeq,\equiv}\circ \cR'= \cR$ rather than
  $\alrel{\simeq,\equiv}\circ \cR'\subseteq\cR$. One can easily come
  up with $\cR'$ which satisfies $\alrel{\simeq,\equiv}\circ
  \cR'\subseteq\cR$ but is not a simulation (take, {\it e.g.}
  $\cR'=\emptyset$). Note also that if $(\cE,\cR)$ is a minimal
  counterexample then the set $\cR(q_0)$ is non-empty for each $q_0\in
  \Q_\cE$. Thus, by taking $\simeq$ to be $\equiv$ and taking
  $X_{[q]_\equiv}$ to be empty for some ${[q]_\equiv}$ above, we will
  ensure that the resulting $\cR'$ is not a simulation and
  $\alrel{\simeq,\equiv}\circ \cR'\subseteq\cR$.  Hence we would have
  declared $\equiv$ to be a good $\equiv$-refinement if we had not
  required the equality!  Thus we may not be able to guarantee any
  progress in the CEGAR loop (see Proposition~\ref{prop:progress}).
\item We require that $(q_\cE, \amdpcomp{q}{\simeq})\in \cR'$. If we
  had not required this condition then one could have achieved a
  ``good'' refinement by just breaking the initial state and taking
  $\cR'$ to be any relation such that $\alrel{\simeq,\equiv}\circ \cR'=
  \cR$ but fails to be a simulation just because $\cR'$ relates
  $q_\cE$ to equivalence classes that does not contain the initial state
  of $\absmdp{\cM}{\simeq}.$
\end{enumerate}
\end{remark}

Suppose $(\cE,\cR)$ is a counterexample for $\absmdp{\cM}{\equiv}$ and
$\psi_S$; hence $\cE\not\sat \psi_S$. Therefore, if $\cR'\subseteq
\Q_\cE\times \amdpcomp{\Q}{\simeq}$ is a canonical simulation then
$(\cE,\cR')$ is a counterexample for $\absmdp{\cM}{\simeq}$ and
$\psi_S$. The following proposition says that if
$\alrel{\simeq,\equiv}\circ \cR'= \cR$ and $(\cE,\cR)$ is invalid for
$(\cM,\equiv)$, then $(\cE,\cR')$ is invalid for $(\cM,\simeq).$ Thus,
a good refinement ensures that at least one of counterexamples
``contained'' within $(\cE,\cR)$ is not a counterexample, and thereby
eliminates at least one spurious counterexample that would not be
eliminated by a bad refinement.

\begin{proposition}
  Let $\cM$ be a MDP with $\Q$ as the set of states, $\simeq$ and
  $\equiv$ be equivalence relations compatible with $\cM$ such that
  $\simeq \subseteq \equiv$. Let $(\cE,\cR)$ be a counterexample for
  $\absmdp{\cM}{\equiv}$ and $\psi_S$ with $\Q_\cE$ as the set of
  states.  Let $\cR'\subseteq \Q_\cE\times \amdpcomp{\Q}{\simeq}$ be a
  canonical simulation such that $\alrel{\simeq,\equiv}\circ \cR'=
  \cR$. Then $(\cE,\cR')$ is a counterexample for
  $\absmdp{\cM}{\simeq}$ and $\psi_S$.  Further, if $(\cE,\cR)$ is
  invalid for $(\cM,\equiv)$ then $(\cE,\cR')$ is invalid for
  $(\cM,\simeq).$
\end{proposition}
\begin{proof}
  That $(\cE,\cR')$ is a a counterexample for $\absmdp{\cM}{\simeq}$
  and $\psi_S$ follows from the fact that $\cR'$ is a simulation and
  that $\cE\not\sat \psi_S$. Assume, by way of contradiction, that
  $(\cE,\cR')$ is valid and consistent with $(\cM,\simeq)$.  Then
  there exists a canonical simulation $\cR_0\subseteq \Q_\cE\times \Q$
  such that $\alrel{\simeq}\circ \cR_0\subseteq \cR'$.  This implies
  that $(\alrel{\simeq,\equiv}\circ\alrel{\simeq})\circ \cR_0\subseteq
  \alrel{\simeq,\equiv}\circ \cR'.$ But the left hand side is
  $\alrel{\equiv} \circ\cR_0$ while the right hand side is $\cR.$ This
  implies that $(\cE,\cR)$ is a valid and consistent with
  $(\cM,\equiv)$ (with $\cR_0$ as the validating simulation).  A
  contradiction!
\end{proof}
 
We conclude this section by showing that good refinements ensure
progress in the CEGAR loop.
\begin{proposition}
\label{prop:progress}
Let $\cM$ be an MDP, and $\simeq$ and $\equiv$ be equivalence
relations compatible with $\cM$ such that $\simeq \subseteq \equiv$. Let
$(\cE,\cR)$ be a counterexample for $\absmdp{\cM}{\equiv}$ and
$\psi_S$. If $\simeq$ is a {\it good $\equiv$-refinement} for
$(\cE,\cR)$, then $\simeq\subsetneq \equiv$.
\end{proposition}
\begin{proof}
  Fix a relation $\cR'\subseteq \amdpcomp{\Q}{\simeq}\times \Q$ such
  that $\alrel{\simeq,\equiv}\circ \cR' =\cR$ but $\cR'$ is not a
  canonical simulation. We now proceed by contradiction. Assume that
  $\simeq=\equiv$. Then $\absmdp{\cM}{\equiv}$ and
  $\absmdp{\cM}{\simeq}$ are the same abstract MDP and
  $\alrel{\simeq,\equiv}$ is the identity relation. Thus $\cR'$ and
  $\cR$ are the same relation.  But $\cR'$ is not a simulation which
  contradicts that fact that $(\cE,\cR)$ is a counterexample for
  $\absmdp{\cM}{\equiv}$ and $\psi_S.$
 \end{proof}

\subsubsection{Algorithm for Refinement}
\label{sec:refine-algo}

In this section, we will show how an abstract model can be refined
based on a spurious counterexample obtained as in Theorem \ref{thm:min}. Before presenting our algorithm,
we recall how the refinement step proceeds for non-probabilistic
systems, through an example. This will help us highlight a couple of
key points about the refinement step in the non-probabilistic case.

\begin{example}\rm
\label{exam:refine-algo}
Recall in the (non-probabilistic) Example \ref{exam:refine}, the
counterexample checking step proceeded from the last state of the
counterexample trace $\cKcexc=\{q_0,q_1,q_3\} \rightarrow
\{q_5,q_6\}\rightarrow \{q_7, q_8\}\rightarrow \{q_{11}\}$ and
confirmed that the concrete state $q_6$ can simulate the path
$\{q_5,q_6\}\rightarrow \{q_7, q_8\}\rightarrow \{q_{11}\}$.  Since
there is no ``concrete'' transition from $\{q_0,q_1,q_3\}$ to $q_6$,
the counterexample checking step concludes that the counterexample is
invalid. The state $\{q_0,q_1,q_3\}$ is the counterpart of
``invalidating abstract state'', the transition
$t=\{q_0,q_1,q_3\}\rightarrow \{q_5, q_6\}$ is the counterpart of
``invaliding transition'', $\{q_6\}$ is counterpart of $R_{old}(\{q_5,
q_6\})$ and $\emptyset$ is counterpart of $R(\{q_0, q_1, q_3\})$.  The
refinement step for non-probabilistic case is obtained by splitting
the equivalence class $\post(t,\{q_0,q_1,q_3\})=\{q_5, q_6\}$ into
$\{q_6\}=R_{old}(\{q_5,q_6\})$ and $\{q_5\}=\{q_5,q_6\}\setminus
R_{old}(\{q_5,q_6\}).$

On the other hand, suppose for the Kripke structure $\cKex$ and its
abstraction $\cKab$, the counterexample
$\cKcexb=\{q_0,q_1,q_3\}\rightarrow\{q_4\}\rightarrow
\{q_{10}\}\rightarrow \{q_{11}\}$ is chosen instead of $\cKcexc$.  In
this case the counterexample generation algorithm declares the
counterexample to be invalid when the initial state of $\cKex$, $q_0$,
fails to be in $R(\{q_0,q_1,q_3\})$ during the counterexample checking
algorithm. In this case, $\{q_0,q_1,q_3\}$ is the ``invalidating''
abstract state; $\{q_0,q_2,q_3\}$ is the counterpart of
$R_{old}(\{q_0,q_1,q_3\})$ and $\{q_3\}$ is the counterpart of
$R(\{q_0,q_1,q_3\})$. For this case, the invalidating abstract state
$\{q_0,q_1,q_3\}$ is itself broken into
$R_{old}(\{q_0,q_1,q_3\})\setminus R(\{q_0,q_1,q_3\})=\{q_0,q_1\}$ and
$\{q_0,q_1,q_3\}\setminus(R_{old}(\{q_0,q_1,q_3\})\setminus
R(\{q_0,q_1,q_3\})=\{q_3\}).$ Thus, in this case the invalidating
abstract state itself is broken into equivalence classes and not its
successor!
\end{example}

Let us examine the refinement step outlined in
Example~\ref{exam:refine-algo} more carefully. There are two cases to
consider: when the invalidating abstract state is not the initial
state, where we only split abstract state that is the target of the
invalidating transition; and when the invalidating abstract state is
the initial state of the counterexample, where we also split the
invalidating abstract state.

To generalize to probabilistic systems, we make the following
observations. Suppose $\absmdp{\cM}{\equiv}$ is the abstraction of
$\cM$ with respect to $\equiv$, and let $(\cE,\rel\inj)$ be a
counterexample for $\absmdp{\cM}{\equiv}$ and $\psi_S$ obtained as in Theorem \ref{thm:min} and which is invalid
for $(\cM,\equiv)$. If $\bar{a}$ is the invalidating abstract state,
and $\mu\in\delta_\cE(\bar{a})$ is the invalidating transition, then
$\post(\mu,\bar{a})$ may contain several states (including $\bar{a}$
itself). Hence, our refinement step will be forced to split several
equivalence classes instead of one as in the case of non-probabilistic
systems. Next, we observe that in the case when the counterexample is
a ``path'' (or more generally a DAG), the algorithm to check validity
only needs to ``process'' each state of the counterexample once.
Hence, if $(R_{old},R)$ is the invalidating witness, then
$R_{old}(\bar{a}) = \gmrel{\equiv}(a)$. Therefore, $R_{old}(\bar{a})
\setminus R(\bar{a})$ is always $\gmrel{\equiv}(a)$ except (possibly)
when the invalidating state $\bar{a}$ is the initial state of the
counterexample. This is the primary reason why for non-probabilistic
systems the invalidating abstract state is never split, except when it
is the initial state. However, when analyzing counterexamples that
could be general MDPs, the counterexample checking algorithm will need
to ``process'' each state multiple times, and then $R_{old}(\bar{a})$
need not be $\gmrel{\equiv}(a)$, at the time the counterexample is
deemed to be invalid. Thus, in our refinement algorithm, we will
be forced to also split the invalidating abstract state.

The above intuitions are formalized in the refinement step shown in
Figure~\ref{fig:refine}; recall that for each $\bar{d}\in \Q_\cE$, we
have $R(\bar{d})\subseteq R_{old}(\bar{d})\subseteq \rel\gamma(d)$. We
conclude by showing that the resulting refinement is a good refinement
(and hence progress is ensured in the CEGAR loop).
\begin{theorem}
  Let $(\cE, \rel\inj)$ be a counterexample, generated using
  Theorem~\ref{thm:min}, for $\absmdp{\cM}{\equiv}$ and safety
  property $\psi_S$, where $\absmdp{\cM}{\equiv}$ is the abstraction
  of $\cM$ with respect to the compatible equivalence relation
  $\equiv$. If the counterexample checking algorithm in
  Figure~\ref{fig:validity} returns (``invalid'', $\bar{a}$, $\mu,
  R_{old}$, $R$), then the refinement $\simeq\subseteq\equiv$ obtained
  as in Figure~\ref{fig:refine} is a good $\equiv$-refinement for
  $(\cE,\rel\inj)$.
\end{theorem}
\begin{proof}
  Let $\cE=(\Q_{\cE},q_{\cE}, \delta_{\cE}, \sL_{\cE})$. Recall that
  $\absmdp{\cM}{\simeq}=(\amdpcomp{\Q}{\simeq},\amdpcomp{q}{\simeq},
  \amdpcomp{\delta}{\simeq}, \amdpcomp{\sL}{\simeq})$ is the 
 
  abstract MDP for $\cM$ and $\simeq$, where the set of states of
  $\absmdp{\cM}{\simeq}$ are equivalence classes under $\simeq$.
  Consider the {\it functional} relation $\cR\subseteq \Q_\cE\times
  \amdpcomp{\Q}{\simeq}$ defined as follows.
\begin{itemize}
\item $(\bar{a}, a')\in \cR$ iff $a'$ is the $\simeq$-equivalence
  class $R_{old}(\bar{a})\setminus R (\bar{a})$.
\item For each $\bar{b} \in \post(\mu,\bar{a})\setminus \bar{a}$,
  $(\bar{b}, b')\in \cR$ iff $b'$ is the $\simeq$-equivalence class
  $R_{old}(\bar{b}).$
\item For each $\bar{c}\in \Q_\cE\setminus ( \post(\mu,\bar{a})\cup
  \bar{a})$, $(\bar{c}, c')\in \cR$ iff $c'=c.$
\end{itemize}
Please note that it is easy to see that by construction
$(q_\cE,\amdpcomp{q}{\simeq})\in \cR$; if $\bar{a}$ is not the initial
state then the observation follows immediately, and otherwise, observe
that $q_\I \in R_{old}(\bar{a})\setminus R(\bar{a})$.  Further, we
clearly have $\alrel{\simeq,\equiv}\circ\cR=\rel\inj$. We shall show
that $\cR$ is not a canonical simulation and hence we can conclude
that $\simeq$ is a good refinement.
 
Now, let $a_0\in\amdpcomp{\Q}{\simeq}$ be the $\simeq$-equivalence
class $\cR_{old}(\bar{a})\setminus R(\bar{a}).$ Observe that
$(\bar{a},a_0)\in \cR$. Next, recall that the violating transition
$\mu \in \delta_{\cE}(\bar{a})$. Hence the desired result will follow
if we can show that for each $\mu_0\in\amdpcomp{\delta}{\simeq}(a_0)$
we have that $\mu \not\simd{\cR} \mu_0.$
 
We proceed by contradiction. Let $\mu'$ be such that
$\mu'\in\amdpcomp{\delta}{\simeq}(a_0)$ and $\mu \simd{\cR} \mu'.$ By
definition of abstractions, there is a $q\in
\gmrel{\simeq}(a_0)=R_{old}(\bar{a})\setminus R(\bar{a})$ and
$\mu_1\in \delta(q)$ such that $\mu'=[\mu_1]_{\simeq}. $ From $\mu
\simd{\cR} \mu'$, we can conclude the following.
\begin{itemize}
\item $\mu(\bar{a}) \leq \mu_1(R_{old}(\bar{a})\setminus
  R(\bar{a}))\leq \mu_1(R_{old}(a)).$
\item For each $\bar{b} \in \post(\mu,\bar{a})\setminus \bar{a}$,
  $\mu(\bar{b}) \leq \mu_1(R_{old}(\bar{b}))$.
\item For $\bar{c}\in \Q_\cE\setminus ( \post(\mu,\bar{a})\cup
  \bar{a})$, $\mu(\bar{c})=0\leq \mu_1(R_{old}(\bar{c})).$
\end{itemize}
For each $\bar{d}\in \Q_\cE$, it follows from construction that
$R_{old}(\bar{d})\subseteq\gmrel{\equiv}({d}).$ Therefore for all
$\bar{d_0},\bar{d_1}\in \Q_\cE$ such that $\bar{d_0}\ne \bar{d_1}$,
$R_{old}(\bar{d_0})\cap R_{old}(\bar{d_1})=\emptyset$.  It now follows
easily from the above observations that $\mu\simd{\cR_{old}}\mu_1$
which contradicts Proposition \ref{prop:violation}.
\end{proof}

\begin{figure}[t]
\begin{center}
\begin{tabular}{|l|}
\hline
\\

The refinement $\simeq$ is obtained from the equivalence $\equiv$ as follows.\\
If $\bar{a}$ is the invalidating abstract state, $\mu \in \delta_\cE(\bar{a})$ the invalidating transition \\
 and  $(R_{old},R)$ the invalidating  witness then:\\
    \hspace{0.3cm} The $\equiv$-equivalence class $\rel\gamma(a)$ is broken into \\
   \hspace{0.7cm}     new $\simeq$-equivalence classes $R_{old}({\bar{a}})\setminus R(\bar{a})$ and  $\rel\gamma(a)\setminus(R_{old}({\bar{a}})\setminus R(\bar{a}))$\\
 \hspace{0.3cm} For each $\bar{b}\in \post(\mu,\bar{a})\setminus \bar{a}$, the $\equiv$-equivalence class $\rel\gamma(b)$ is broken into\\
   \hspace{0.7cm}               new $\simeq$-equivalence classes             $R_{old}({\bar{b}})$ and  $\rel\gamma(b)\setminus R_{old}({\bar{b}})$\\
\hspace{0.3cm} No other $\equiv$-equivalence class is refined.\\
\\
\hline
\end{tabular}
\end{center}
\caption{Refinement algorithm based on invalid counterexamples}
\label{fig:refine}
\end{figure}

\subsection{Counter-example checking for weak safety}
\label{sec:onthefly}

In this section we outline an algorithm that given a counterexample
$\cE$ for an MDP $\cM$ and weak safety property $\psi_{WS}$ either
determines that $\cE$ is not a valid counterexample, or finds a finite
unrolling of $\cE$ that is simulated by $\cM$, and witnesses the fact
that $\cM$ does not satisfy $\psi_{WS}$. The algorithm unrolls $\cE$
on the fly, and does not construct the unrolled MDP explicitly. The
running time of the algorithm depends on the height of the unrolling,
which if small, can result in faster checking than the algorithm shown
in Figure~\ref{fig:validity}. Before presenting the algorithm, we
introduce some notation that we will find useful in describing the
algorithm. Recall that given an MDP $\cM$, a state $q$ of $\cM$ and
$k\in \Nats$ the $k$-th unrolling of MDP rooted at $q$ is denoted by
$\cM^q_k.$

\begin{notation} 
  Given $k\in \Nats$ and states $q,q' \in \cM$, we say that $q
  \preceq_k q'$ if $\cM^q_{k}\preceq \cM^{q'}_k$. Given a PCTL formula
  $\psi$ we say that $q\sat_k\psi$ if $\cM^q_{k}\sat\psi$.
\end{notation}

We observe the following two facts.  If $q\preceq q'$ then $q
\preceq_k q'$ for all $k$.  The proof of
Theorem~\ref{thm:tree-cntr-strict-liveness} implies that for a weak
safety formula $\psi_{WS}$ if $q \not \sat \psi_{WS}$, then there is a
$k_0$ s.t. $q\not\sat_{k_0} \psi_{WS}$. These two facts can be
combined to obtain a counter-example checking algorithm for the
weak-safety fragment of PCTL as we shall describe shortly.

For the rest of the Section, we fix the following notation.
$\cM=(\Q,q_\I, \delta, \sL)$ is the (original) MDP that we are
checking against weak safety property $\psi_{WS}$ and $\equiv$ is an
equivalence relation that is compatible with $\cM$. Assuming
$\absmdp{\cM}{\equiv}$ violates the safety property $\psi_{WS}$, we
will denote the minimal counterexample obtained as in Theorem
\ref{thm:min} by $(\cE, \rel\inj)$. Let $\cE=
(\Q_\cE,q_\cE,\delta_\cE,\sL_\cE)$ . For a state $a=[q]_\equiv$ in
abstraction $\absmdp{\cM}{\equiv}$, $\gmrel{\equiv}(a)={\{q'\,|\,
  q\equiv q'\}}$ is the concretization map.  The relation
$\{(\bar{a},q)\; |\; q \in \gmrel{\equiv}(a), \bar{a}\in \Q_\cE\}$ shall
be denoted by $R_\I$.  Finally, we shall use $\psi_{SL}$ to denote the
strict liveness formula obtained by negating $\psi_{WS}$.
$\SubForm(\psi_{SL})$ will denote the set of {\PCTL} strict liveness
subformulas of $\psi_{SL}$ and $\PathForm(\psi_{SL})$ will denote the
set of path subformulas of $\psi_{SL}$\footnote{A path subformula of a
  PCTL-liveness formula $\psi_L$ are formulas of the kind $\X\psi_L$
  and $\psi_L\cU\psi_L$ }.
  
The proposed algorithm iteratively constructs the relations
$R_k=R_\I\, \cap \preceq_k$ and $\Sat_k=\{(\bar{a},\psi) \mbar
\bar{a}\sat_k \psi,\bar{a} \in Q_\cE,\psi \in \SubForm(\psi_{SL}) \}$.
We make the following observations.
\begin{enumerate}
\item If the set $R_k(\bar{a})=\{q \mbar (\bar{a},q)\in R_k\}$ becomes
  $\emptyset$ for some $k$ and $\bar{a} \in \Q_\cE$, then
  $(\cE,\rel\inj)$ is invalid for $(\cM,\equiv)$. We can also call the
  counterexample invalid if the initial concrete state $\ic$ is not
  contained in $R_k(q_\cE)$.
\item If $\Sat_k(q_\cE,\psi_{SL})$ and $\ic\in R_k(q_\cE)$ then $\ic
  \sat_k \psi_{SL}$ also.  Thus, the concrete MDP violates
  the given safety property and we can report this.
\end{enumerate}
This iteration must end as a consequence of Theorem
\ref{thm:tree-cntr-strict-liveness}. The computation also needs to
compute the function $\MaxProb_k$ defined as follows. Given
$\bar{a}\in Q_\cE$ and a path formula $\phi\in\PathForm(\psi_{SL})$,
$\MaxProb_k$ gives the maximum probability (over all schedulers) of
$\phi$ being true in $\cE^{\bar{a}}_{k}$. The relations $R_{k+1}$,
$\Sat_{k+1}$ and $\MaxProb_{k+1}$ can be computed using $R_{k}$,
$\Sat_k$, and $\MaxProb_k$ and do not need other previous values.

\begin{figure}[!htp]

\begin{tabular}{|l|}
\hline
\\

Initially \\
\hspace{0.3cm} $R_\curr=\{(\bar{a},q)\; |\; q \in \rel\gamma(\bar{a}), \bar{a} \in \Q_\cE\}$\\
\hspace{0.3cm} $\Sat_\curr=\{(\bar{a},\psi)\; |\; \bar{a}\sat_0\psi,\psi\in\SubForm(\psi_{SL}), \bar{a} \in \Q_\cE\}$\\

\hspace{0.3cm}  $\MaxProb_\curr[\bar{a}][\phi]= \MaxProb_0(\bar{a})(\phi)$ for $\bar{a} \in \Q_\cE,\phi \in \PathForm(\psi_{SL})$ \\

While (true) \\
\hspace{0.3cm}do \\
  \hspace{1cm} If $\ic \notin R_\curr(q_\cE) $ return ``Counter-example not simulated''\\
  \hspace{1cm} If $\Sat(q_\cE,\psi_{SL})$ return ``Safety Violated''\\
  \hspace{1cm} for each  $\bar{a} \in \Q_\cE$\\
  \hspace{1.2cm} do \\
                         
    \hspace{1.5cm} $R_{\tmp, \bar{a}}= \{q\mbar (\bar{a},q) \in R_\curr \textrm{ and } \forall
    \mu \in \delta_\cE(\bar{a})\exists \mu' \in \delta(q).\ \mu
    \simd{R_\curr} \mu'\}$\\
    
    \hspace{1.5cm} If $R_{\tmp, \bar{a}}=\emptyset$ return ``Counter-example not simulated'' \\
    \hspace{1.5cm} COMPUTE($\bar{a},\Sat_\curr, \MaxProb_\curr, \Sat_{\tmp, \bar{a}}, \MaxProb_{\tmp, \bar{a}}$) \\
  \hspace{1.2cm} od\\

  \hspace{1cm} $\Sat_\curr=\{(\bar{a},\psi)\mbar \psi\in \Sat_{\tmp, \bar{a}}\}$\\
  \hspace{1cm} $R_\curr=\{(\bar{a},q)\mbar q\in R_{\tmp, \bar{a}}\}$\\

  \hspace{1cm} $\MaxProb_\curr[\bar{a}][\phi]=\MaxProb_{\tmp, \bar{a}}[\phi] $ for each $a\in \Q_\cE, \phi \in \PathForm(\psi)$\\
\hspace{0.3cm} od\\
\\

The procedure COMPUTE returns $\Sat_{\tmp, \bar{a}}$, the set of sub-formulas of 
$\psi_{SL}$ satisfied by $\bar{a}$\\ 
in the ``next'' unrolling of the tree with root $\bar{a}$. It also
returns  $\MaxProb_{\tmp, \bar{a}}$,\\
that given a path formula $\phi$ gives the 
maximum probability of $\phi$ being true in \\
the next unrolling. COMPUTE is defined as follows.  
\\
\\
COMPUTE($\bar{a},\Sat_\curr,  \MaxProb_\curr, \Sat_{\tmp, \bar{a}},\MaxProb_{\tmp, \bar{a}}$) \\
\hspace{0.3cm}Fix an enumeration $\varphi_1,\ldots,\varphi_n$ of  the set 
$\{\varphi\; |\; \varphi \in \PathForm(\psi_{SL})\cup$ \\ 
\hspace{0.8cm}$\SubForm(\psi_{SL})\}$ 
such that $size(\varphi_i) \leq size(\varphi_j)$ for $i\leq j$.\\

\hspace{0.3cm} Initially\\
\hspace{0.5cm} $\Sat_{\tmp, \bar{a}} =\emptyset$\\
\hspace{0.5cm} $\MaxProb_{\tmp, \bar{a}}[\phi]=0$ for all $\phi \in \PathForm(\psi_{SL}).$\\

\hspace{0.5cm} For  $i= 1 $ to  $n$ \\

\hspace{0.8cm} If $\varphi_i$ is $p$ and  $p\in \sL_\cE(\bar{a})$ then
   $\Sat_{\tmp, \bar{a}}=\Sat_{\tmp, \bar{a}}\cup \{p\}$\\
 
\hspace{0.8cm} If $\varphi_i$ is $\neg p$ and $p\notin \sL_\cE(\bar{a})$ then
   $\Sat_{\tmp, \bar{a}}=\Sat_{\tmp, \bar{a}}\cup \{\neg p\}$\\

\hspace{0.8cm} If $\varphi_i$ is $\psi_1 \disj \psi_2$ and ($\psi_1\in \Sat_{\tmp, \bar{a}}$ or $\psi_2\in \Sat_{\tmp, \bar{a}}$) then\\
\hspace{1.2cm}
                            $\Sat_{\tmp, \bar{a}}=\Sat_{\tmp, \bar{a}}\cup \{\varphi_i\}$  \\
 
\hspace{0.8cm} If $\varphi_i$ is $\psi_1 \conj \psi_2$, $\psi_1\in \Sat_{\tmp, \bar{a}}$ and $\psi_2\in \Sat_{\tmp, \bar{a}}$  then\\
\hspace{1.2cm}  
                            $\Sat_{\tmp, \bar{a}}=\Sat_{\tmp, \bar{a}}\cup \{\varphi_i\}$  \\
                           
\hspace{0.8cm} If $\varphi_i$ is $\neg\cP_{\leq p}(\phi)$ and $\MaxProb_{\tmp, \bar{a}}[\phi] >p$ \\
\hspace{1.2cm}then $\Sat_{\tmp, \bar{a}}=\Sat_{\tmp, \bar{a}}\cup \{\varphi_i\}$  \\
 \\
 
 \hspace{0.8cm} If $\varphi_i$ is $\X\psi$ then\\
 \hspace{1.2cm}   $\MaxProb_{\tmp, \bar{a}}[\varphi_i]=\max_{\mu \in \delta_\cE(\bar{a})}\mu(\{\bar{b}\; |\; (\bar{b},\psi) \in \Sat_\curr\})$ \\

  \hspace{0.8cm} If $\varphi_i$ is $\psi_1\cU\psi_2$ then\\
 \hspace{1.2cm}   If $\psi_2\in \Sat_{\tmp, \bar{a}}$ then  $\MaxProb_{\tmp, \bar{a}}[\varphi_i]=1$\\
  \hspace{1.2cm}  else\\
  \hspace{1.6cm}  If $\psi_1\notin  \Sat_{\tmp, \bar{a}}$ then  $\MaxProb_{\tmp, \bar{a}}[\varphi_i]=0$\\
  \hspace{1.6cm} else $\MaxProb_{\tmp, \bar{a}}[\varphi_i]= \max_{\mu \in \delta_\cE(\bar{a})} \sum_{\bar{b}\in \Q_\cE} (\mu(\bar{b}))\;(\MaxProb_\curr[\psi_1\cU\psi_2][\bar{b}])$\\ 
                                                            
\\
\hline
\\             
\end{tabular}
\caption{On the fly algorithm for checking Strict Liveness}
\label{fig:onthefly}
\end{figure}
 
Figure~\ref{fig:onthefly} gives the details of this algorithm. At the
beginning of $(k+1)$-th unrolling of the while loop, the relations
$R_\curr$ and $\Sat_\curr$ are the relations $R_k$ and $\Sat_k$
respectively. The (doubly-indexed) array
$\MaxProb_\curr[\bar{a}][\psi]$ is the function
$\MaxProb_k[\bar{a}][\psi]$ . Within the while loop,
$R_{\tmp,\bar{a}}$ is the set $R_{k+1}(\bar{a})$ while
$\Sat_{\tmp,\bar{a}}$ is the set $\{\psi\mbar \bar{a}\sat_{k+1} \psi
\in \SubForm(\psi_{SL})\}$, and $\MaxProb_{\tmp,\bar{a}}[\psi]$ is
$\MaxProb_{k+1}[\bar{a}][\psi]$. $R_\curr$, $\Sat_\curr$,
and $\MaxProb_\curr$ are updated {\it after}
$R_{\tmp,\bar{a}}$, $\Sat_{\tmp,\bar{a}}$ and
$\MaxProb_{\tmp,\bar{a}}$ are computed for {\it all} states
$\bar{a}\in \Q_\cE.$ The following proposition follows easily from the
observations made in the Section.
\begin{proposition} 
  The algorithm in Figure~\ref{fig:onthefly} terminates.  If the
  algorithm returns ``Counter-example not simulated'' then the
  counterexample obtained using Theorem \ref{thm:min} is not valid. If
  the algorithm returns ``Safety Violated'' then $\cM\not\sat
  \psi_{WS}.$
\end{proposition}
 
Finally, we observe that the algorithm in Figure~\ref{fig:onthefly}
may be made more efficient in practice as follows. First, since we are
dealing with strict liveness fragment, the sequence $\Sat_k$ is an
increasing sequence and the function $\MaxProb_k[\bar{a}][\psi]\leq
\MaxProb_{k+1}[\bar{a}][\psi]$. Hence, only needs to compute
$\Sat_{k+1}\setminus \Sat_{k}$ and $\MaxProb_{k+1}-\MaxProb_{k}$. This
optimization shall be explored in future work.

\section{Related Work}
\label{sec:related-work}

\noindent
{\bf Abstraction Schemes:}
Abstractions have been extensively studied in the context of
probabilistic systems. General issues in defining good abstractions as
well as specific proposals for families of abstract models are
presented
in~\cite{jl91,hut04-1,nor04,hut05,Dajjl01,Dajjl02,flw06,kklw07,mon05,knp06-1,im}.
Recently, theorem-prover based algorithms for constructing
abstractions of probabilistic systems based on predicates have been
presented~\cite{wzh07}. Another notion that has been recently proposed
is the notion of a ``magnifying-lens
abstraction''~\cite{magnifying-lens}, which can be used to assist in
the model checking process, by approximating the measure of the
satisfaction of path formulas for sets of concrete states; the method
is not an abstraction in the traditional sense in that neither is an
abstract model explicitly constructed, nor is the model used for
reasoning, one that simulates the concrete model.

\vspace*{0.2cm}
\noindent
{\bf Counterexamples:} 
The notion of counterexamples is critical for the approach of
counterexample guided abstraction refinement. Criteria for defining
counterexamples are identified in~\cite{cjlv02}, along with a notion
of counterexamples for branching-time properties and non-probabilistic
systems. The problem of defining counterexamples for probabilistic
systems has received considerable attention recently.  Starting from
the seminal papers~\cite{jazzar-counterex,hk07-1}, the notion of sets
of paths with high measure as counterexamples has been used for DTMCs,
CTMCs, and MDPs~\cite{hk07-1,hk07-2,jazzar}. Another definition that
has been proposed is that of DTMCs (or purely probabilistic models)
in~\cite{chatterjee-games,holmanns}. Our notion of counterexample is
different from these proposals and we demonstrate that these other
proposals are not rich enough for the class of properties we consider.

\vspace*{0.2cm}
\noindent
{\bf Automatic Abstraction-Refinement:} 
In the context of probabilistic systems, automatic abstract-refinement
was first considered in~\cite{Dajjl01,Dajjl02}. There are two main
differences with our work. First, they consider only reachability
properties. Second, the refinement process outlined
in~\cite{Dajjl01,Dajjl02} is not counterexample based, but rather
based on partition refinement.  What this means is that their
refinement is biased towards separating states that are not bisimilar,
rather than states that are ``distinguished'' by the property, and so
it is likely that their method refines more than needed.

Counterexample guided refinement has been used as the basis of
synthesizing winning strategies for 2-player stochastic games
in~\cite{chatterjee-games}. Though the problem 2-player games is more
general than verification, the specific model considered
by~\cite{chatterjee-games} has some peculiarities and so does not
subsume the problem or its solution presented here. First, in their
model, states are partitioned into ``non-deterministic'' states that
have purely nondeterministic transitions and ``probabilistic'' states
that have purely probabilistic transitions.  The abstraction does not
abstract any of the ``probabilistic states''; only the
``non-deterministic'' states are collapsed. This results in larger
abstract models, and obviates certain issues in counterexample
checking that we deal with. Second, they take counterexamples to be
finite models without nondeterminism. They can do this because they
consider a simpler class of properties than we do, and as we show in
Section~\ref{sec:no-dtmc}, DTMCs are not rich enough for all of
safe-\PCTL . Next, their counterexample checking algorithm is
different than even the one used in the context of non-probabilistic
systems. They consider a counterexample to be valid only if {\it all} the
concrete states corresponding to the abstract states in the
counterexample can simulate the behavior captured by the
counterexample. Thus, they deem certain counterexamples to be
spurious, even if they will be recognized as providing enough evidence
for the violation of the property by other CEGAR schemes (including
ours). Finally, they do not have a precise statement characterizing
the qualities of their refinement algorithm.

In a recent paper,~\cite{holmanns} consider CEGAR for probabilistic
systems. They consider very special types of reachability properties
namely, those that can be expressed by formulas of the form $\cP_{\leq
  p}(\psi_1\cU\psi_2)$ where $\psi_1$ and $\psi_2$ are propositions
(or boolean combinations of propositions). For this class of
properties, they use DTMCs as the notion of counterexamples : the
counterexample obtained is a pair $(\cS,{\cM}^\cS)$ where $\cS$ is a
memoryless scheduler. As we show, DTMCs cannot serve as counterexample
for the richer class of properties considered here. For the
counterexample checking algorithm, they generate a finite set,
$\mathsf{ap}$, of {\it abstract} paths of $\absmdp{\cM}{\equiv}^\cS$
in decreasing order of measures such that the total measure of these
paths is $>p$. Then, they build (on-the-fly) a ``concrete'' scheduler
which maximizes the  measure of the paths in $\mathsf{ap}$ that are
simulated by the original MDP.  Let $\mathsf{p}_\mathsf{total}$ be the
maximum probability (under all schedulers) of $\psi_1\cU \psi_2$ being
satisfied by the abstract MDP, $\mathsf{p}_\mathsf{ap}$ be the total
measure of $\mathsf{ap}$ and $\mathsf{p}_\mathsf{max}$ be the maximum
measure of ``abstract'' paths in $\mathsf{ap}$ simulated by the
concrete MDP.  If $\mathsf{p}_\mathsf{max}>p$ then the counterexample
is declared to be valid and if
$\mathsf{p}_\mathsf{total}-\mathsf{p}_\mathsf{ap}+\mathsf{p}_\mathsf{max}\leq
p$ then the counterexample is declared to be invalid and the
abstraction refined.  If neither is the case, then \cite{holmanns}
heuristically decide either to generate more abstract paths or to
refine the abstraction. The refinement is based upon refining some
``spurious'' abstract path (namely, a path that is not simulated by
the concrete system).
There is, however, no formal statement
characterizing progress based on the refinement algorithm outlined
in~\cite{holmanns}. 


\section{Conclusions and future work}

We presented a CEGAR framework for MDPs, where an MDP $\cM$ is
abstracted by another MDP $\cA$ defined using an equivalence on the
states of $\cM$. Our main contributions when presenting this framework
were a definition for the notion of a counterexample, along with
algorithms to compute counterexamples, check their validity and
perform automatic refinement based on an invalid counterexample.

There are a number of interesting questions left open for future
investigation. First these ideas need to be implemented and
experimented with. In order for this approach to be scalable, symbolic
algorithms for a lot of the steps outlined here, will be required.
Next, when constructing minimal counterexamples, the order in which
transitions are considered for elimination, crucially affects the
final size of the counterexample. Good heuristics for ordering
transitions to obtain small counterexamples, must be identified.

{\bf Acknowledgments.}  The authors thank the following people:
Chandra Chekuri for many fruitful discussions especially relating to
the algorithmic aspects of finding counterexamples and checking
validity of counterexamples; Radha Jagadeesan for discussions on
simulations and the safety fragment of PCTL; anonymous referees for
sending pointers to~\cite{jazzar,holmanns} and for encouraging us to
formally articulate the guarantees of our refinement algorithm.

\bibliography{references} \bibliographystyle{acmtrans}

\begin{thebibliography}{}

\bibitem[\protect\citeauthoryear{Aljazzar, Hermanns, and Leue}{Aljazzar
  et~al\mbox{.}}{2005}]{jazzar-counterex}
{\sc Aljazzar, H.}, {\sc Hermanns, H.}, {\sc and} {\sc Leue, S.} 2005.
\newblock Counterexamples for timed probabilistic reachability.
\newblock In {\em Proceedings of International Conference on Formal Modelling
  and Analysis of Timed Systems}. 177--195.

\bibitem[\protect\citeauthoryear{Aljazzar and Leue}{Aljazzar and
  Leue}{2007}]{jazzar}
{\sc Aljazzar, H.} {\sc and} {\sc Leue, S.} 2007.
\newblock Counterexamples for model checking of markov decision processes.
\newblock Tech. Rep. soft-08-01, University of Konstanz.

\bibitem[\protect\citeauthoryear{Baier, Engelen, and Majster-Cederbaum}{Baier
  et~al\mbox{.}}{2000}]{bem00}
{\sc Baier, C.}, {\sc Engelen, B.}, {\sc and} {\sc Majster-Cederbaum, M.~E.}
  2000.
\newblock Deciding bisimularity and simularity for probabilistic processes.
\newblock {\em Journal of Computer and System Sciences\/}~{\em 60,\/}~1,
  187--231.

\bibitem[\protect\citeauthoryear{Baier, Katoen, Hermanns, and Wolf}{Baier
  et~al\mbox{.}}{2005}]{bkhw05}
{\sc Baier, C.}, {\sc Katoen, J.-P.}, {\sc Hermanns, H.}, {\sc and} {\sc Wolf,
  V.} 2005.
\newblock Comparative branching-time semantics for markov chains.
\newblock {\em Information and Computation\/}~{\em 200}, 149--214.

\bibitem[\protect\citeauthoryear{Bianco and de~Alfaro}{Bianco and
  de~Alfaro}{1995}]{Luca}
{\sc Bianco, A.} {\sc and} {\sc de~Alfaro, L.} 1995.
\newblock Model checking of probabalistic and nondeterministic systems.
\newblock In {\em Proceedings of the International Conference on the
  Foundations of Software Technology and Theoretical Computer Science}.
  499--513.

\bibitem[\protect\citeauthoryear{Chatterjee, Henzinger, Jhala, and
  Majumdar}{Chatterjee et~al\mbox{.}}{2005}]{chatterjee-games}
{\sc Chatterjee, K.}, {\sc Henzinger, T.}, {\sc Jhala, R.}, {\sc and} {\sc
  Majumdar, R.} 2005.
\newblock Counter-example {G}uided {P}lanning.
\newblock In {\em Proceedings of Uncertainity in Artificial Intelligence}.
  104--111.

\bibitem[\protect\citeauthoryear{Clarke, Grumberg, Jha, Lu, and Veith}{Clarke
  et~al\mbox{.}}{2000}]{cgjlv00}
{\sc Clarke, E.}, {\sc Grumberg, O.}, {\sc Jha, S.}, {\sc Lu, Y.}, {\sc and}
  {\sc Veith, H.} 2000.
\newblock Counterexample-guided abstraction refinement.
\newblock In {\em Proceedings of the International Conference on Computer Aided
  Verification}. 154--169.

\bibitem[\protect\citeauthoryear{Clarke, Jha, Lu, and Veith}{Clarke
  et~al\mbox{.}}{2002}]{cjlv02}
{\sc Clarke, E.}, {\sc Jha, S.}, {\sc Lu, Y.}, {\sc and} {\sc Veith, H.} 2002.
\newblock Tree-like counterexamples in model checking.
\newblock In {\em Proceedings of the IEEE Symposium on Logic in Computer
  Science}. 19--29.

\bibitem[\protect\citeauthoryear{D'Argenio, Jeannet, Jensen, and
  Larsen}{D'Argenio et~al\mbox{.}}{2001}]{Dajjl01}
{\sc D'Argenio, P.~R.}, {\sc Jeannet, B.}, {\sc Jensen, H.~E.}, {\sc and} {\sc
  Larsen, K.~G.} 2001.
\newblock Reachability analysis of probabilistic systems by successive
  refinements.
\newblock In {\em Proceedings of the International Workshop on Performance
  Modeling and Verification}. 39--56.

\bibitem[\protect\citeauthoryear{D'Argenio, Jeannet, Jensen, and
  Larsen}{D'Argenio et~al\mbox{.}}{2002}]{Dajjl02}
{\sc D'Argenio, P.~R.}, {\sc Jeannet, B.}, {\sc Jensen, H.~E.}, {\sc and} {\sc
  Larsen, K.~G.} 2002.
\newblock Reduction and refinement strategies for probabilistic analysis.
\newblock In {\em Proceedings of the International Workshop on Performance
  Modeling and Verification}. 57--76.

\bibitem[\protect\citeauthoryear{de~Alfaro and Roy}{de~Alfaro and
  Roy}{2007}]{magnifying-lens}
{\sc de~Alfaro, L.} {\sc and} {\sc Roy, P.} 2007.
\newblock Magnifying-lens abstraction for {M}arkov {D}ecision {P}rocesses.
\newblock In {\em Proceedings of the International Conference on Computer Aided
  Verification}. 325--338.

\bibitem[\protect\citeauthoryear{Desharnais}{Desharnais}{1999a}]{des}
{\sc Desharnais, J.} 1999a.
\newblock Labelled markov processes.
\newblock Ph.D. thesis.

\bibitem[\protect\citeauthoryear{Desharnais}{Desharnais}{1999b}]{des99}
{\sc Desharnais, J.} 1999b.
\newblock Logical characterisation of simulation for markov chains.
\newblock Tech. Rep. CSR-99-8, University of Birmingham.
\newblock Proceedings of the International Workshop on Performance Modeling and
  Verification.

\bibitem[\protect\citeauthoryear{Desharnais, Gupta, Jagadeesan, and
  Panangaden}{Desharnais et~al\mbox{.}}{2000}]{radha}
{\sc Desharnais, J.}, {\sc Gupta, V.}, {\sc Jagadeesan, R.}, {\sc and} {\sc
  Panangaden, P.} 2000.
\newblock Approximating labeled {M}arkov processes.
\newblock In {\em Proceedings of the IEEE Symposium on Logic in Computer
  Science}. 95--106.

\bibitem[\protect\citeauthoryear{Dodis and Khanna}{Dodis and
  Khanna}{1999}]{dk99}
{\sc Dodis, Y.} {\sc and} {\sc Khanna, S.} 1999.
\newblock Designing networks with bounded pairwise distance.
\newblock In {\em Proceedings of the ACM Symposium on the Theory of Computing}.
  750--759.

\bibitem[\protect\citeauthoryear{Fecher, Leucker, and Wolf}{Fecher
  et~al\mbox{.}}{2006}]{flw06}
{\sc Fecher, H.}, {\sc Leucker, M.}, {\sc and} {\sc Wolf, V.} 2006.
\newblock Don't know in probabilistic systems.
\newblock In {\em Proceedings of the International SPIN Workshop on Model
  Checking Software}. 71--88.

\bibitem[\protect\citeauthoryear{Garey and Johnson}{Garey and
  Johnson}{1979}]{garey}
{\sc Garey, M.} {\sc and} {\sc Johnson, D.} 1979.
\newblock {\em Computers and {I}ntractability: {A} {G}uide to the {T}heory of
  {NP}-{C}ompleteness}.
\newblock Freeman.

\bibitem[\protect\citeauthoryear{Han and Katoen}{Han and
  Katoen}{2007a}]{hk07-1}
{\sc Han, T.} {\sc and} {\sc Katoen, J.-P.} 2007a.
\newblock Counterexamples in probabilistic model checking.
\newblock In {\em Proceedings of the International Conference on Tools and
  Algorithms for the Construction and Analysis of Systems}. 72--86.

\bibitem[\protect\citeauthoryear{Han and Katoen}{Han and
  Katoen}{2007b}]{hk07-2}
{\sc Han, T.} {\sc and} {\sc Katoen, J.-P.} 2007b.
\newblock Providing evidence of likely being on time --- {C}ounterexample
  generation for {CTMC}.
\newblock In {\em Proceedings of the International Symposium on Automated
  Technology for Verification and Analysis}.

\bibitem[\protect\citeauthoryear{Hermanns, Wachter, and Zhang}{Hermanns
  et~al\mbox{.}}{2008}]{holmanns}
{\sc Hermanns, H.}, {\sc Wachter, B.}, {\sc and} {\sc Zhang, L.} 2008.
\newblock Probabilistic {CEGAR}.
\newblock In {\em Proceedings of the International Conference on Computer Aided
  Verification}.
\newblock Also available as technical report number 33, AVACS, University of
  Saarland, 2007.

\bibitem[\protect\citeauthoryear{Huth}{Huth}{2004}]{hut04-1}
{\sc Huth, M.} 2004.
\newblock An abstraction framework for mixed non-deterministic and
  probabilistic systems.
\newblock In {\em Validation of Stochastic Systems: {A} Guide to Current
  Research}. 419--444.

\bibitem[\protect\citeauthoryear{Huth}{Huth}{2005}]{hut05}
{\sc Huth, M.} 2005.
\newblock On finite-state approximants for probabilistic computation tree
  logic.
\newblock {\em TCS\/}~{\em 346}, 113--134.

\bibitem[\protect\citeauthoryear{Jonsson and Larsen}{Jonsson and
  Larsen}{1991}]{jl91}
{\sc Jonsson, B.} {\sc and} {\sc Larsen, K.~G.} 1991.
\newblock Specification and refinement of probabilistic processes.
\newblock In {\em Proceedings of the IEEE Symposium on Logic in Computer
  Science}. 266--277.

\bibitem[\protect\citeauthoryear{Katoen, Klink, Leucker, and Wolf}{Katoen
  et~al\mbox{.}}{2007}]{kklw07}
{\sc Katoen, J.-P.}, {\sc Klink, D.}, {\sc Leucker, M.}, {\sc and} {\sc Wolf,
  V.} 2007.
\newblock Three-valued abstraction for continuous-time markov chains.
\newblock In {\em Proceedings of the International Conference on Computer Aided
  Verification}. 311--324.

\bibitem[\protect\citeauthoryear{Kwiatkowska, Norman, and Parker}{Kwiatkowska
  et~al\mbox{.}}{2006}]{knp06-1}
{\sc Kwiatkowska, M.}, {\sc Norman, G.}, {\sc and} {\sc Parker, D.} 2006.
\newblock Game-based abstraction for markov decision processes.
\newblock In {\em Proceedings of the International Conference on Quantitative
  Evaluation of Systems}. 157--166.

\bibitem[\protect\citeauthoryear{McIver and Morgan}{McIver and
  Morgan}{2004}]{im}
{\sc McIver, A.} {\sc and} {\sc Morgan, C.} 2004.
\newblock {\em Abstraction, {R}efinement and {P}roof for {P}robabilistic
  {S}ystems}.
\newblock Springer.

\bibitem[\protect\citeauthoryear{Monniaux}{Monniaux}{2005}]{mon05}
{\sc Monniaux, D.} 2005.
\newblock Abstract interpretation of programs as markov decision processes.
\newblock {\em Science of Computer Programming\/}~{\em 58}, 179--205.

\bibitem[\protect\citeauthoryear{Norman}{Norman}{2004}]{nor04}
{\sc Norman, G.} 2004.
\newblock Analyzing randomized distributed algorithms.
\newblock In {\em Validation of Stochastic Systems: {A} Guide to Current
  Research}. 384--418.

\bibitem[\protect\citeauthoryear{Rutten, Kwiatkowska, Norman, and
  Parker}{Rutten et~al\mbox{.}}{2004}]{marta-book}
{\sc Rutten, J.~M.}, {\sc Kwiatkowska, M.}, {\sc Norman, G.}, {\sc and} {\sc
  Parker, D.} 2004.
\newblock {\em Mathematical Techniques for Analyzing Concurrent and
  Probabilistic Systems}.
\newblock AMS.

\bibitem[\protect\citeauthoryear{Segala}{Segala}{2006}]{segala}
{\sc Segala, R.} 2006.
\newblock Probability and nondeterminism in operational models of concurrency.
\newblock In {\em Proceedings of the International Conference on Concurrency
  Theory}. 64--78.

\bibitem[\protect\citeauthoryear{Segala and Lynch}{Segala and
  Lynch}{1994}]{segalalynch}
{\sc Segala, R.} {\sc and} {\sc Lynch, N.~A.} 1994.
\newblock Probabilistic simulations for probabilistic processes.
\newblock In {\em Proceedings of the International Conference on Concurrency
  Theory}. 481--496.

\bibitem[\protect\citeauthoryear{Wachter, Zhang, and Hermanns}{Wachter
  et~al\mbox{.}}{2007}]{wzh07}
{\sc Wachter, B.}, {\sc Zhang, L.}, {\sc and} {\sc Hermanns, H.} 2007.
\newblock Probabilistic model checking modulo theories.
\newblock In {\em Proceedings of the International Conference on Quantitative
  Evaluation of Systems}.

\bibitem[\protect\citeauthoryear{Zhang, Hermanns, Eisenbrand, and Jansen}{Zhang
  et~al\mbox{.}}{2007}]{zhej07}
{\sc Zhang, L.}, {\sc Hermanns, H.}, {\sc Eisenbrand, F.}, {\sc and} {\sc
  Jansen, D.~N.} 2007.
\newblock Flow faster: {E}fficient decision algorithms for probabilistic
  simulation.
\newblock In {\em Proceedings of the International Conference on Tools and
  Algorithms for the Construction and Analysis of Systems}. 155--170.

\end{thebibliography}
\end{document}